\documentclass[preprint]{aastex}

\def\simgr{\,\hbox{\hbox{$ > $}\kern -0.8em \lower 1.0ex\hbox{$\sim$}}\,}
\def\simle{\,\hbox{\hbox{$ < $}\kern -0.8em \lower 1.0ex\hbox{$\sim$}}\,}

\shortauthors{Halpern \& Thorstensen}
\shorttitle{X-ray Selected Cataclysmic Binaries}

\def\int{({\it INTEGRAL\/)}/IBIS}
\def\chandra{{\it Chandra}}
\def\xmm{{\it XMM--Newton}}
\def\sw{{\it Swift\/}}

\newcommand{\pbcOhThreeTwoFive}{PBC J0325.6$-$0820}
\newcommand{\swiftOhFiveOhThree}{Swift J0503.7$-$2819}
\newcommand{\swiftOhFiveTwoFive}{Swift J0525.6+2416}
\newcommand{\swiftOhSixOneFour}{Swift J0614.0+1709}
\newcommand{\swiftOhSixTwoThree}{Swift J0623.9$-$0939}
\newcommand{\pbcOhSevenOhSix}{PBC J0706.7+0327}
\newcommand{\swiftOhSevenOneSeven}{Swift J0717.8$-$2156}
\newcommand{\swiftOhSevenFourNine}{Swift J0749.7$-$3218}
\newcommand{\swiftOhEightTwoOh}{Swift J0820.6$-$2805}
\newcommand{\swiftOhNineThreeNine}{Swift J0939.7$-$3224}
\newcommand{\pbcOneSevenFourOh}{4PBC J1740.7+0603}
\newcommand{\swiftTwoOneTwoFour}{Swift J2124.6+0500}
\newcommand{\swiftTwoThreeFourOne}{Swift J2341.0+7645}
\newcommand{\swiftTwoOhOneFive}{Swift J2015.9+3715}
\newcommand{\pbcTwoOhOneFive}{4PBC J2015.5+3711}
\newcommand{\rxj}{RX J2015.6+3711}

\begin{document}

\title{Optical Studies of Thirteen Hard X-ray Selected 
Cataclysmic Binaries from the \sw-BAT Survey\footnote
{Based on observations obtained at the MDM Observatory,
operated by Dartmouth College, Columbia University, 
Ohio State University, Ohio University, and the University of Michigan.}
}

\author{Jules P. Halpern}
\affil{Columbia Astrophysics Laboratory,
Columbia University,
550 West 120th Street, 
New York, NY 10027, USA}

\author{John R. Thorstensen}
\affil{Department of Physics and Astronomy, Dartmouth College,
Hanover NH, 03755, USA}

\begin{abstract}
From a set of thirteen cataclysmic binaries that were discovered
in the \sw\ Burst Alert Telescope (BAT) survey, we conducted
time-resolved optical spectroscopy and/or time-series photometry of
eleven, with the goal of measuring their orbital periods and searching
for spin periods. 
Seven of the objects in this study are new optical identifications.
Orbital periods are found for
seven targets, ranging from 81 minutes to 20.4 hours.
\pbcOhSevenOhSix\ is an AM Herculis star (polar)
based on its emission-line variations and large
amplitude photometric modulation on the same period.
\swiftTwoThreeFourOne\ may be a polar, although the
evidence here is less secure.
Coherent pulsations are detected from two objects,
\swiftOhFiveOhThree\ (975~s) and
\swiftOhSixOneFour\ (1412~s and 1530~s,
spin and beat periods, respectively),
indicating that they are probable intermediate polars
(DQ Herculis stars).  For two other stars, longer spin
periods are tentatively suggested.
We also present the discovery of a 2.00 hour X-ray modulation from
\rxj, possibly a contributor to \swiftTwoOhOneFive, and likely a polar.
\end{abstract}


\section{INTRODUCTION}

Cataclysmic variables (CVs) are accreting binaries in which
a dwarf star donates mass to a white dwarf (WD) via Roche-lobe
overflow.  CVs such as dwarf novae are often discovered
in outburst by large-scale optical  time-domain surveys.
Ultraviolet and X-ray surveys
find CVs in all states of luminosity.  Hard X-ray energies
($>15$~keV) preferentially select systems in which the
magnetic field of the WD is strong enough to truncate the
accretion disk at a magnetospheric boundary, or prevent it
from forming completely.  In these systems, an accretion
column is channeled onto the magnetic pole(s), where thermal
bremsstrahlung X-rays are radiated from a shock just
above the surface of the WD.

In polars (AM Herculis stars), the magnetic field is strong
enough to channel matter directly from the companion, and
the WD rotation is locked to the binary orbit,
or nearly so in the few asynchronous polars. 
Polars are also characterized by their optical circular
polarization, and optical/IR humps in their spectra
from cyclotron radiation in the strong magnetic field.
Intermediate polars (IPs, or DQ Herculis stars)
have weaker magnetic fields and a truncated accretion
disk.  The spin period of the WD in an IP is detected
as a coherent oscillation in X-ray or optical emission
from a rotating hot spot, at a shorter period
than the orbital period of the binary.  Sometimes the beat
period between the spin and orbit periods is seen,
due to reprocessed emission.

We have been studying newly discovered CVs and CV candidates
from the hard X-ray surveys of the {\it International Gamma-Ray 
Astrophysics Laboratory} \int, and the \sw\ Burst Alert Telescope
(BAT), using time-resolved optical spectroscopy to measure their
orbital periods, and time-series photometry to search
for spin and/or orbital modulation.
The Swift-BAT 70 month hard X-ray survey \citep{bau13}
lists 55 CVs, of which 41 are magnetic: 31 IPs and 10 polars.
IPs outnumber polars in hard X-ray surveys,
presumably because of their higher accretion rates, and also
because in polars the accretion may be more ``blobby,'' depositing
some fraction of the energy directly onto the surface of the
WD, which is radiated at temperatures of tens of eV.

In Thorstensen \& Halpern (2013, hereafter Paper~I), we presented
data and findings on 10 hard X-ray selected CVs, three of which
were unidentified X-ray sources prior to our work.
The present paper continues this effort
with information on another thirteen objects.
In particular, we now concentrate on the unclassified \sw-BAT sources,
of which there are 88, including the 23 indicated
simply as ``Galactic'' based on their coordinates.

The targets were selected based on their accessibility
from the northern hemisphere,
and properties likely to favor magnetic CVs.  Spectral
information from the literature was consulted.  Unidentified
\sw-BAT sources, for which we examined pointed observations
with the \sw\ X-ray Telescope (XRT) and UV/Optical Telescope
(UVOT), were chosen if they had bright,
variable counterparts in the XRT and/or a relatively
bright and/or blue optical counterpart in the UVOT.  These
criteria were found to be uniformly successful in identifying
CVs.  In addition, we identified several non-variable X-ray
sources with fainter optical counterparts as active
galactic nuclei, which will be reported elsewhere.
This paper identifies seven new CV counterparts, as well as
characterizing six that were previously identified
spectroscopically but lack detailed studies in the literature.
We also present in an Appendix an X-ray result on a CV
that may have been detected in hard X-rays.

\section{EQUIPMENT AND TECHNIQUES}
\label{sec:techniques}

Our instrumentation, and reduction and analysis
procedures are essentially identical to those described
in Paper~I, and are summarized in this section.
All of our optical data are from the MDM Observatory, which
comprises the 1.3~m McGraw-Hill telescope and the
2.4~m Hiltner telescope, both on the southwest ridge
of Kitt Peak, Arizona.  With a single exception,
the radial velocity studies to search for the orbital
periods were done on the 2.4~m, while high-cadence
photometry sensitive to spin periods was carried
out on the 1.3~m.
 
\subsection{Spectroscopy}

All of our radial velocity studies used the
modular spectrograph, as described in
Paper~I.  In addition to wavelength calibrations using
comparison lamps of Hg, Ne, and Xe taken during twilight, 
we used the night-sky \ion{O}{1} $\lambda5577$ line 
to track and correct for spectrograph flexure
during the night.  Most of our velocities are from the 
the 2.4m telescope, where we rotated the 
instrument to orient the slit close to the parallactic
angle when at large hour angles and zenith distances.
Although the $1^{\prime\prime}$
projected slit width does not permit accurate spectrophotometry,
flux standards were applied to the spectra, with
an expected accuracy of $\sim 20\%$ for the 
averages spectra suggested by experience.
Some spectra of \swiftTwoOneTwoFour, and all 
the data we used for \swiftOhNineThreeNine, are from
the McGraw-Hill 1.3m telescope, again with the modular
spectrograph.  At the 1.3m telescope, we left the
spectrograph slit oriented north-south for all
observations.  

We reduced our spectra using Pyraf scripts that for
the most part called
IRAF\footnote {IRAF is distributed by the National Optical Astronomy
Observatory, which is operated by the Association of
Universities for Research in Astronomy (AURA) under
cooperative agreement with the National Science
Foundation.} tasks. 
To extract one-dimensional spectra, we implemented
the algorithm published by \citet{horne}.   We
measured emission-line radial velocities using
convolution algorithms described by \citet{sy80} and
\citet{shafter83}.  To measure synthetic 
magnitudes from our spectra we used the
\citet{bessell} tabulation of the $V$ passband
and the IRAF {\tt sbands} task.

To search for periods in the spectroscopic
time series, we used a ``residual-gram'' algorithm
\citep{tpst}.  Observing constraints sometimes led
to ambiguities in the daily cycle count.  To resolve them,
we used the Monte Carlo method of \citet{tf85}, which
employs the ``discriminatory power'' statistic,
which is the fraction of Monte Carlo trials
that result in the correct period being chosen, rather than
an alias, by virtue of having the smallest residuals.

For four newly identified objects we have only single
spectra that were obtained on two observing runs on
the 2.4m.  These used the Boller and Chivens CCD spectrograph
(CCDS) and the Ohio State Multi-Object Spectrograph (OSMOS).
Descriptions of these instruments can be found on the MDM
Observatory web 
page\footnote{http://mdm.kpno.noao.edu/index/Instrumentation.html}. 

\subsection{Time-Series Photometry}
 
Most of the time-series photometry used ``Templeton'', 
a thinned, back-illuminated SITe CCD with
$1024\times1024$ $24 \mu$ pixels, each subtending
$0.\!^{\prime\prime}508$ at the 1.3~m.
To minimize read time, the readout was usually
windowed and binned $2\times2$.  Exposures ranged from
10--30~s, with 3~s dead-time between exposures.
For one star, we used the thermoelectrically cooled
Andor Ikon DU-937N CCD camera as described in Paper~I.
It has a dead-time of only 11.92~ms
when binned $4\times4$ to give a pixel size of
$1.\!^{\prime\prime}1$.  Filters used were $V$ for bright stars or,
more commonly, broadband Schott BG38 or GG420 to maximize
throughput for fainter stars.   BG38 passes a broad band
from roughly 3250 to 6500~\AA, with a significant red leak.
GG420 is a long-pass filter transmitting $\lambda>4200$~\AA\
that is useful for suppressing scattered moonlight.

Differential photometry with respect to a single
comparison star that was typically 2--4 mag brighter
than the variable was performed using the IRAF task
{\tt phot}. Approximate magnitudes were
derived for the BG38 filter by averaging the $B$ and $R$
magnitudes of the comparison star in the USNO B1.0 catalog
\citep{mon03}, or adopting the $R$ magnitude of the
comparison star in the GG420 filter.
Sequences on individual stars ranged from 1.9--9.6 hours.
Period searches used a standard power-spectrum analysis on
these evenly sampled light curves.

\section{RESULTS ON INDIVIDUAL OBJECTS}
\label{sec:individuals}

The objects observed are listed in Table~\ref{tab:objects},
which provides accurate celestial coordinates,
approximate magnitudes, and spin periods where discovered.
Figure~\ref{fig:charts} gives finding charts of the objects
for which we have our own direct images.
Table~\ref{tab:velocities} lists the radial velocity data,
and Table~\ref{tab:parameters} gives parameters of the
best-fit sinusoids.
Figures~\ref{fig:montage1} and \ref{fig:montage2} show mean
spectra, radial velocity periodograms, and folded radial
velocities for seven of the objects.  Time-series photometry,
and additional identification spectra appear in subsequent
figures.

\subsection{\pbcOhThreeTwoFive} 
\label{sec:pbc0325}

This X-ray source is listed in the second Palermo \sw-BAT
hard X-ray catalog \citep{cus10b}, and was identified
spectroscopically as a CV by \citet{par14}.
It is also known as 1RXS J032540.0$-$081442.

The mean spectrum (Figure~\ref{fig:montage1}) shows Balmer and
\ion{He}{1} emission lines, and \ion{He}{2} $\lambda 4686$ less than 
half the strength of H$\beta$.  The lines are relatively
narrow, with the FHWM of H$\alpha$ being about 10 \AA ,
or 460 km s$^{-1}$.  An M-type absorption spectrum
is evident.  This is probably from the secondary, but because
our signal-to-noise is insufficient to measure absorption
radial velocities, it could be from an interloper or tertiary star.
We could not classify the secondary accurately, but it is 
roughly consistent with a spectral type between M2 and M5.
Decompositions with earlier types
tended to oversubtract the bandhead near 6150~\AA\ 
and undersubtract the bands in the red; later types 
tended to oversubtract the far red end of the
spectrum when the bands were matched

We have spectra from two observing runs, in 2014 January and
2014 October.  The radial velocities are consistent with 
two different periods, 0.0872(1) d and 0.0857(1) d, which are
separated by one cycle per five days.  The number of 
cycles in interval between the observing runs is unknown,
but the periods near 0.0872 d are constrained to be 
$$P = {263.324 \pm 0.005\ \hbox{d} \over 3020 \pm 10},$$
where the denominator is an integer.
The radial velocity
amplitude in January was $112\pm12$~km~s$^{-1}$, but
increased to $178\pm7$~km~s$^{-1}$ in October, so in
Figure~\ref{fig:montage1} we distinguish the two runs' velocities
with different symbols.  The physical significance of the
change is unclear.

We also obtained time-series photometry of \pbcOhThreeTwoFive\
on three consecutive nights in 2013 December.  The light curves
in Figure~\ref{fig:pbc0325} show a wave of amplitude $\lesssim 0.1$~mag
that produces a best period of $125.0\pm0.5$ minutes ($0.0868\pm0.0003$~d)
in the combined power spectrum.  This is consistent with 
and favors the larger of the two alternative spectroscopic periods.

As noted above, the M-dwarf contribution could
not be constrained precisely, but we used the Monte Carlo
procedure described in Paper I to loosely constrain the 
system's distance, on the assumption that the light does
arise from the secondary.  Choosing the spectral 
type from a uniform distribution between M2 to M6, and 
adopting liberal uncertainties
for the secondary's mass and flux contribution,
yielded a distance of 400 ($+170, -160$) pc.
Most CVs in this period range do not show an
obvious secondary star in their spectrum, so this is 
likely to be a rather low-luminosity system. 

\subsection{\swiftOhFiveOhThree}
\label{sec:swift0503}

We selected \swiftOhFiveOhThree\ from the 70-month \sw-BAT all-sky
hard X-ray survey \citep{bau13}, and identified its counterpart
as a bright, variable X-ray source in \sw\ XRT images of the field,
and as a UV bright star in the \sw\ UVOT.
This object is also known as 1RXS J050350.6$-$282324 and
1WGA J0503.8$-$2823.

The mean spectrum of \swiftOhFiveOhThree\ (Figure~\ref{fig:montage1})
shows strong H, \ion{He}{1}, and \ion{He}{2} lines
on a blue continuum, with \ion{He}{2} $\lambda$4686 roughly equal in strength to
H$\beta$.  The H$\alpha$ line has an equivalent width (EW) of $\approx55$ \AA 
\footnote{In this paper, we assign positive equivalent widths to
emission features.}.  
A search for radial-velocity periodicities in H$\alpha$ was inconclusive, but the 
\ion{He}{1} and \ion{He}{2} lines showed a stronger modulation, so we adopted 
an average of the \ion{He}{1} $\lambda$5876 and \ion{He}{2} $\lambda$4686 velocities
for the period analysis.  They showed a strong periodicity near 81 minutes, 
but with ambiguity in the daily cycle count.  Figure~\ref{fig:greyscales} shows
a phase-binned greyscale representation of the spectra at the 81-minute
period; the strong velocity modulation is especially clear in 
\ion{He}{2} $\lambda$4686 and $\lambda$5411.  However, the lines do not show
the asymmetric, shifting wings typical of an AM Her star, or polar.  

We obtained time-series photometry of \swiftOhFiveOhThree\
on five nights in 2013 December and 2014 January, as shown
in Figure~\ref{fig:swift0503}.  A power spectrum of the
combined data shows three clear peaks, at $4896\pm4$~s,
$2449\pm1$~s, and $975.2\pm0.2$~s.
The first two signals are undoubtedly the orbital period and its
harmonic, corresponding to $P_{\rm orb}=0.05667\pm0.00005$~d,
or 81.60(7) min,
which agrees with the favored spectroscopic value in Table~3,
and resolves the ambiguity in the spectroscopic cycle count.
The 975.2~s signal could be the spin period of the WD,
which would make \swiftOhFiveOhThree\ another rare example of
an IP below the (orbital) period gap.

\subsection{\swiftOhFiveTwoFive}
\label{sec:swift0525}

Also known as 1RXS J052523.2+241331, the spectroscopy
of \swiftOhFiveTwoFive\ showed \ion{He}{2} $\lambda4686$
stronger than H$\beta$, suggesting a magnetic CV identification
\citep{tor07,masetti12}. \citet{ber15} discovered a 226.3~s
X-ray oscillation in \xmm\ data from this source,
establishing it as an IP.
No other variability was detected in the 8.8~hr X-ray observation.
In addition, \citet{ber15} rejected previous 
evidence for an optical eclipse in \swiftOhFiveTwoFive\ 
\citep{ram09} as being due to a telescope tracking problem.

We have obtained 20.5 hours of time-series photometry on
\swiftOhFiveTwoFive\ over four nights in 2013 and 2014.  The
best of these optical light curves, using a $V$ or BG38 filter,
are shown in Figure~\ref{fig:swift0525}.  The longest is 9.6~hr
and contains no eclipse.  While showing variability
on a variety of time scales between $\sim500$~s
and $\sim6$~hr, the optical data do not display
the secure 226.3~s X-ray period, which should be detected easily
if its amplitude were the same as in soft X-rays, $\approx10\%$.
Figure~\ref{fig:swift0525} shows the power spectrum of the 9.6~hr
BG38 time series, in which any signal at 226.3~s is within
the noise, and has a semi-amplitude of $\lesssim0.004$~mag, or a
pulsed fraction of $\lesssim0.37\%$.
\citet{ber15} concluded from its energy dependence that the
X-ray pulse modulation is due to photoelectric absorption
in the preshock-accretion flow.  Possibly most of the visible
light is coming from a larger region.

\subsection{\swiftOhSixOneFour}
\label{sec:swift0614}

We identified this source from \citet{bau13} as the
the brightest one in XRT images of the field.
Only one optical spectrum of \swiftOhSixOneFour\ was obtained,
in the blue region, which displays higher order Balmer lines and
\ion{He}{2} $\lambda4686$ comparable in strength to H$\beta$
(Figure~\ref{fig:spectra}a), identifying it as a probable
magnetic CV.

Time-series photometry on three consecutive nights (Figure~\ref{fig:swift0614})
shows a clear pair of beating periods, $1529.8(9)$~s and $1412.3$(9)~s.
In addition, the difference frequency between these appears in the
power spectrum as a peak at $5.00(5)$~hr.  We interpret the 1412~s
period as the spin period and the 5.00~hr period as the
orbital period.  (Examining the power spectrum in Figure~\ref{fig:swift0614},
1-day aliases of these values are an alternative possibility:
1390~s and 4.15~hr.)
The 1530~s period is the strongest signal, which
corresponds to the beat between the spin and orbit frequencies.
When seen in X-rays in intermediate polars, this phenomenon is
attributed to diskless accretion, or else to part of the
accretion stream passing over the disk and coupling
directly to the magnetosphere \citep{hel93}.  The beat period
is then due to reprocessing of the spin signal by material fixed
in the orbiting frame of the system.

\subsection{\swiftOhSixTwoThree}
\label{sec:swift0623}

We identified this source from \citet{bau13} as the
the brightest one in XRT images of the field.
It corresponds to 1RXS J062406.9$0-$093815.
Only one optical spectrum of \swiftOhSixTwoThree\ was obtained
(Figure~\ref{fig:spectra}b), which shows Balmer lines, \ion{He}{1}
and \ion{He}{2} $\lambda4686$, identifying it as a CV.  A single
5.3~hr time series on this bright, $V\approx14$ object,
obtained during partly cloudy weather, shows flickering typical
of a CV, but no obvious period.  Additional data are
needed to characterize this system.

\subsection{\pbcOhSevenOhSix}
\label{sec:pbc0706}

This X-ray source is listed in the second Palermo \sw-BAT
hard X-ray catalog \citep{cus10b}, and also in \citet{bau13}
as Swift J0706.8+0325.  It was identified
spectroscopically as a CV by \citet{par14}.
It is also known as 1RXS J070648.8+032450.

The PPMXL catalog \citep{ppmxl} gives a proper motion of
$[\mu_X, \mu_Y] = [-68, -43]$ mas yr$^{-1}$ for the 
$\approx17$ mag optical counterpart.  At 100 pc, this corresponds to 
a transverse velocity of 40 km s$^{-1}$, which suggests
that the distance is less than a few hundred pc.

The mean spectrum (Figure~\ref{fig:montage1})
shows strong, relatively narrow lines, with 
broad wings.  \ion{He}{2} $\lambda$4686 is present but weaker than 
H$\beta$.  The H$\alpha$ emission velocities define an 
unambiguous period near 102 minutes, with a very large
amplitude of 300 km s$^{-1}$.  The line profiles in the 
phase-averaged greyscale (Figure~\ref{fig:greyscales}) show 
a narrow, low-amplitude component and a more diffuse,
high-amplitude component nearly in antiphase.  This is
typical of polars, in which accretion 
occurs through a magnetically-channeled funnel onto
a synchronously-rotating WD, without an accretion
disk. 

We obtained a single night of time-series photometry on
\pbcOhSevenOhSix\ (Figure~\ref{fig:pbc0706}), which supports
the AM Her classification.  It shows two large-amplitude spikes
per rotation, such that the peak in the power spectrum
appears at half the orbital period.  The spiky, flaring
nature of this light curve resembles that of the remarkable
and enigmatic IGR J19552+0044 (Paper I), which is likely to
be a polar, or perhaps an asynchronous polar \citep{ber13}.
In \pbcOhSevenOhSix, the spectroscopic and photometric
periods are identical to within the $\sim2\%$ uncertainty on the latter.

\subsection{\swiftOhSevenOneSeven\ and \swiftOhSevenFourNine}
\label{sec:swift0717}

Both of these objects from \citet{bau13} were identified as
the brightest sources in XRT images of their respective fields,
and as faint blue objects in the UVOT.  The first corresponds to
1RXS J071748.9$-$215306.
Spectra were obtained showing H$\alpha$ and H$\beta$ emission,
as well as faint lines of \ion{He}{1}, identifying them
as likely CVs (Figure~\ref{fig:spectra}c,d).
We do not have any time-series data on
\swiftOhSevenOneSeven\ or \swiftOhSevenFourNine,
which would be needed to further characterize their nature.

\subsection{\swiftOhEightTwoOh}
\label{sec:swift0820}

We selected \swiftOhEightTwoOh\ from \citet{bau13},
and identified its counterpart as a bright,
variable X-ray source in \sw\ XRT images of the field,
and as a UV bright star in the \sw\ UVOT.
This object is also known as 1RXS J082033.6$-$280457.
It was also listed in \citet{cus10b}, and was identified
as a CV by \citet{par14}, who showed a spectrum having
prominent lines of H, \ion{He}{1}, and \ion{He}{2}.

We were only able to obtain the brief light curves
displayed in Figure~\ref{fig:swift0820}.  Although a
possible peak in the power spectrum is present
at $2485\pm50$~s on one night, we regard this as
only a candidate for a spin period
because the light curve covers only five cycles
of the oscillation, so it could just be flickering.
Thus, the IP (DQ~Her)
classification is tentative.

\subsection{\swiftOhNineThreeNine}
\label{sec:swift0939}

This source is associated with 1RXS~J093949.2$-$322620
in \citet{bau13}.  We identified it from XRT and UVOT images.
Nearly all our spectral data on \swiftOhNineThreeNine\
are from the 1.3m telescope in 2015 March. 
The mean spectrum (Figure~\ref{fig:montage1})
shows a strong, blue continuum with relatively weak 
emission lines -- the emission EW of H$\alpha$ is $\approx 22$ \AA\ -- 
suggesting that the accretion disk is optically
thick and the luminosity relatively high.

The H$\alpha$ radial velocities are modulated with an 8.5-hour
period.  Because of hour-angle constraints at the
very southerly declination, the choice of 
daily cycle count is not entirely unambiguous, but the
velocity modulation is defined well enough that the 
Monte Carlo test from \citet{tf85} gives a discriminatory
power of over 95\% and a correctness likelihood
near unity.  Despite the long orbital period, we see
no clear sign of a secondary star in the mean spectrum,
which corroborates the suggestion that the luminosity
is relatively high.

Our time-series photometry (Figure~\ref{fig:swift0939}) shows
a possible period of 2670(7)~s, but the short runs do
not cover many of its cycles.  Thus, we consider
\swiftOhNineThreeNine\ only a candidate for a DQ~Her classification,
requiring further observations, similar to the case of
\swiftOhEightTwoOh.

\subsection{\pbcOneSevenFourOh}
\label{sec:pbc1740}

\pbcOneSevenFourOh\ does not yet appear in any published paper, but
we identified it from \sw\ XRT and UVOT images taken
on 2014 November 4 and 5.  It is included
in the fourth Palermo \sw-BAT hard X-ray 
catalog\footnote{http://bat.ifc.inaf.it/100m\_bat\_catalog/100m\_bat\_catalog\_v0.0.htm}.
We were only able to obtain brief time series on \pbcOneSevenFourOh\
due to clouds and technical problems.  They show typical flickering
of CVs, but no obvious periods (Figure~\ref{fig:pbc1740}).

The mean spectrum (Figure~\ref{fig:montage2})
shows a blue continuum with strong, broad, 
single-peaked emission lines -- the EW and FWHM of H$\alpha$ 
are 125~\AA\ and 1400 km~s$^{-1}$, respectively. \ion{He}{2} $\lambda$4686 
is about 1/3 the strength of H$\beta$.  A period search of 
the H$\alpha$ velocities was inconclusive, but 
the \ion{He}{1} $\lambda$5876 and $\lambda$6678 lines 
showed a low-amplitude velocity modulation near 101 minutes, 
with daily aliasing.  The modulations in the two
lines independently gave the same period and phase, within 
uncertainties, so the periodicity is likely to be real.
However, the low signal-to-noise yielded rather poor
frequency discrimination, so the daily cycle count 
remains ambiguous.  Figure~\ref{fig:greyscales} shows a 
phase-averaged greyscale representation of the data.
The $\sim 30$ km s$^{-1}$ velocity modulation would
not be apparent at this scale, but unfortunately 
other subtle phenomena (line wings, S-waves, etc.)
do not appear; the absence of such corroboration suggests that
the period determination should be viewed with some caution.

\subsection{\swiftTwoOneTwoFour}
\label{sec:swift2124}

We selected this X-ray source from \citet{bau13}.
It is also listed in \citet{cus10b} as PBC J2124.5+0503,
and was identified spectroscopically as a CV by \citet{par14}.
The mean spectrum (Figure~\ref{fig:montage2})
shows a strong, blue continuum
with relatively weak emission lines, characteristic
of a novalike variable.  The EW 
of H$\alpha$ is only $17$~\AA , and its FWHM is 
650 km~s$^{-1}$. The \ion{He}{1} lines are rather weak,
but \ion{He}{2} $\lambda4686$ is strong, and the 
\ion{C}{3}/\ion{N}{3} blend at $\lambda4640$ is about half the strength
of H$\beta$.  On the basis of the latter line features,
and its bright continuum, \citet{hal13}
originally suggested an LMXB classification for
\swiftTwoOneTwoFour.  However, its high Galactic latitude
($b=-30.\!^{\circ}5$) and absence of brighter historical
X-ray detections favor a CV interpretation.

As is often the case with novalike variables, the radial 
velocity modulation is ragged, despite ample
counting statistics.  However, our data do show
a significant modulation at 20.1 hours, which 
is consistent across many observing runs.  Daily aliases
are possible but unlikely.  The cycle count between
observing runs is not determined well, with several
reasonable fits spaced in frequency by approximately one cycle
per 57~days.  

\subsection{\swiftTwoThreeFourOne}
\label{sec:swift2341}

\swiftTwoThreeFourOne\ was identified spectroscopically as a CV by 
\citet{lut12}; it corresponds to 1RXS J234015.8+764207.
The mean spectrum (Figure~\ref{fig:montage2})
shows strong Balmer, \ion{He}{1}, and \ion{He}{2}
emission lines, with \ion{He}{2} $\lambda$4686 (which is near the edge
of the  spectral range, these data having been taken with a smaller
detector than the others) comparable in strength to H$\beta$.  
The averaged line profiles have a narrow core and broad wings.
The H$\alpha$ radial velocities show a clear periodicity at
3.7 hours.  The line
core velocities (measured using convolution with a narrow Gaussian)
and the line wing velocities (measured using Gaussians separated by 
$\approx1500$ km s$^{-1}$ give the same period, within the 
uncertainties; however, the line core velocities lag behind
the wings by about 0.2 cycles.  For the cores, the 
velocity semi-amplitude $K$ is $77 \pm 8$ km s$^{-1}$, while
for the wings it is $151 \pm 13$ km s$^{-1}$. 
The data in Table~\ref{tab:parameters} and {Figure~\ref{fig:montage2}
are for the line wings.  \swiftTwoThreeFourOne\
may be a polar, but the case for this is not as strong
as it is for \pbcOhSevenOhSix.  However, we also
obtained time-series photometry of \swiftTwoThreeFourOne\ on
three consecutive nights using the Andor camera on the 1.3m telescope.
Figure~\ref{fig:swift2341} shows a highly modulated signal at the
spectroscopic period, and no shorter periods,
which supports classification as a polar.

\section{CONCLUSIONS}
\label{sec:conclusions}

We identified seven new CV counterparts of \sw-BAT survey sources,
and studied an additional six that were previously known.
Two of the new identifications, \swiftOhFiveOhThree\ (975~s)
and \swiftOhSixOneFour\ (1412~s), are IPs based on likely spin periods
detected in their photometry. The latter also shows
a strong signal at 1530~s, the beat period between the
spin and the 5.00~hr orbit.
Another two, \swiftOhEightTwoOh\ and \swiftOhNineThreeNine\
have possible spin periods that need to be confirmed,
2585~s and 2670~s, respectively.  The previously known
source \pbcOhSevenOhSix\ is shown to be a polar based
on its two-component line profiles, high radial
velocity amplitude, and large photometric
variability on the same 102~minute period.
Its unusual light curve resembles that of the
enigmatic IGR J19552+0044 (Paper I).
\swiftTwoThreeFourOne\ may be a polar based on its
spectra and orbital light curve.
Polarimetry could clarify its nature.

Notably, our long optical light curve of \swiftOhFiveTwoFive\
does not detect the 226~s spin period
discovered in X-rays by \citet{ber15}, to a sensitive limit.  
On the other hand,
\citet{ber15} detect in X-rays the 1218~s optical spin period
of 1RXS J045707.4+452751 (Swift J0457.1+4528) reported in
Paper~I, at a similar amplitude.

Although the small subset of \sw-BAT CVs studied here
is not a well-defined ``sample'' in any quantitative sense,
these results are consistent with the expectation
that hard X-ray selection favors magnetic CVs, and
that IPs outnumber polars in hard X-ray surveys
because of their higher accretion rates, and because
in polars some fraction of the accretion energy
is radiated as a soft X-ray blackbody from the WD surface.

We obtained orbital periods for seven objects
from radial-velocity spectroscopy and time-series
photometry, the latter useful to resolve
ambiguities in cycle count.  These range from
81 minutes to 20.4 hours.
With an orbital period of 81 minutes,
\swiftOhFiveOhThree\ can be added to the small but
growing number of IPs with orbital periods below
the period gap, similar to IGR J18173$-$2509 (92 minutes)
and AX J1853.3$-$012 (87 minutes) from Paper I.
\swiftOhFiveOhThree\ also has the unusually large value of
$P_{\rm spin}/P_{\rm orb}=0.2$ and, similar to IGR J18173$-$2509,
is another exception to the observation of \citet{sca10}
that all hard X-ray-detected IPs have $P_{\rm spin}/P_{\rm orb}<0.1$.

\acknowledgments

We thank Sean McGraw for obtaining the identification spectra
of \swiftOhSixOneFour\ and \swiftOhSixTwoThree.
We gratefully acknowledge support from NSF grant
AST--1008217. 

\section{APPENDIX\\AN X-RAY PERIOD FROM \rxj}

\rxj\ was spectroscopically identified as a CV by \citet{hal01,hal02}
during a survey of the field of the $\gamma$-ray source 3EG J2016+3657.
It was flagged as likely a magnetic CV because its \ion{He}{2}
$\lambda4686$ emission line is comparable in strength to H$\beta$.
It is in a region crowded with high-energy sources from soft
X-rays to TeV $\gamma$-rays, including
the hard X-ray source IGR J20159+3713, which corresponds to
the BAT source \swiftTwoOhOneFive/\pbcTwoOhOneFive,
and the supernova remnant/pulsar wind nebula CTB 87 = G74.9+1.2.
As discussed in detail by \citet{bas14}, the identification
of the source of the hard X-rays in this region
is ambiguous, principally because there is also
the blazar B2013+370 only $1.\!^{\prime}7$ from
\rxj, and they are both within the hard X-ray error circle.
\citet{bau13} identify the BAT source as the blazar while
\citet{cus10b} identify it as the CV; neither is obviously wrong.
\citet{bas14} favor the blazar as the primary source above
20~keV, while considering that the CV probably makes some contribution.

We cannot resolve this ambiguity; however we report here the
discovery of a 2.00~hour period in the X-ray emission from \rxj,
detected near the edge of an archival \chandra\ ACIS-I
image of CTB 87 (ObsID 11092, \citealt{mat13}).
With a continuous exposure of 70~ks and a time resolution of 3.2~s,
this observation is well suited to searching for an orbital or spin
period.  Figure~\ref{fig:rxjperiod} shows the light curve and power
spectrum in the 0.5--2~keV band, chosen for display because the amplitude
of modulation is greatest in soft X-rays.  A period of 7215(31)~s
is clearly detected, as well as significant power at 3596(9)~s,
the harmonic.  The origin of the harmonic power is evident in
the asymmetric light curve.

The complete set of energy-resolved folded light curves from 0.5-8~keV
are shown in Figure~\ref{fig:rxjfold}.
The disappearance of the modulation in the harder X-rays indicates
that its origin is photoelectric absorption. This effect also rules
out an instrumental origin, for example, the dithering of the satellite.
(The dither periods are 1000~s and 707~s in perpendicular directions;
neither is commensurate with the measured 2.00~hours.)  The complex
shape and large amplitude of the soft X-ray light curve suggests an
AM~Her origin, in which the modulation is due to the varying view of
the accretion stream. The absence of any shorter period in X-rays,
such as would be expected from the spin of an IP,
is consistent with this conclusion.  Unfortunately, we do not
yet have enough optical time-series data on \rxj\ for comparison.

\clearpage

\begin{deluxetable}{lcccccccc}
\tablecolumns{9}
\tablewidth{0pt}
\tablecaption{Stars Observed}
\tablehead{
\colhead{Name} &
\colhead{R.A.\tablenotemark{a}} &
\colhead{Decl.\tablenotemark{a}} &
\colhead{$V$} & 
\colhead{Ref\tablenotemark{b}} & 
\colhead{Data\tablenotemark{c}} & 
\colhead{Class\tablenotemark{d}} &
\colhead{$P$} &
\colhead{Ref\tablenotemark{e}}\\
\colhead{} &
\colhead{(h\ \ \ m\ \ \ s)} &
\colhead{($^\circ$\ \ \ $'$\ \ \ $''$)} & 
\colhead{} & 
\colhead{} & 
\colhead{} & 
\colhead{} &
\colhead{(s)} &
\colhead{}
}
\startdata
\pbcOhThreeTwoFive & 03 25 39.43 & $-$08 14 42.8 & 18.3 & S & S, T & N \\  
\swiftOhFiveOhThree & 05 03 49.25 & $-$28 23 08.8 & 18.1 & S & I, S, T & DQ & 975.2(2) & 1 \\  
\swiftOhFiveTwoFive & 05 25 22.75 & +24 13 33.5 & 16.6 & D & S, T & DQ & 226.28(7) & 2 \\
\swiftOhSixOneFour & 06 14 12.28 & +17 04 32.6 & 17.5 & D & I, T & DQ & 1412.3(9)\tablenotemark{f} & 1 \\
\swiftOhSixTwoThree & 06 24 06.18 & $-$09 38 52.1 & 14.1 & A & I, T &  & \\
\pbcOhSevenOhSix & 07 06 48.94 & +03 24 47.5 & 18.1 & S & S, T & AM \\ 
\swiftOhSevenOneSeven & 07 17 48.25 & $-$21 53 01.8 & 18.9 & S & I & \\ 
\swiftOhSevenFourNine & 07 49 31.99 & $-$32 15 37.1 & 18.5 & B & I & & \\
\swiftOhEightTwoOh & 08 20 34.10 & $-$28 04 58.7 & 17.8 & D & T & DQ? & 2485(50)? & 1 \\
\swiftOhNineThreeNine & 09 39 49.65 & $-$32 26 22.1 & 17.3 & S & I, S, T & DQ? & 2670(7)? & 1\\ 
\pbcOneSevenFourOh & 17 40 45.86 & +06 03 51.1 & 15.9 & S & I, S, T & \\ 
\rxj & 20 15 36.96 & +37 11 23.2 & 17.8 & S & X & AM & 7215(31) & 3 \\
\swiftTwoOneTwoFour & 21 24 12.44 & +05 02 43.6 & 12.7 & S & S & N \\ 
\swiftTwoThreeFourOne & 23 40 20.65 & +76 42 10.5 & 17.9 & S & S, T & AM? & \\ 
\enddata
\tablenotetext{a}{Coordinates are for J2000.0 (ICRS), either
from the PPMXL catalogue \citep{ppmxl}, the Sloan Digital
Sky Survey \citep{sdssdr12}, or derived from astrometric fits to our own images.
Estimated uncertainty is $\pm 0.\!^{\prime\prime}2$.}
\tablenotetext{b}{Source of approximate $V$ magnitude: S = our
spectrophotometry; D = our direct image; A = APASS \citep{apass}, 
as listed in the UCAC4 \citep{ucac4}; 
B = interpolated from Schmidt-plate magnitudes in USNO B1.0 \citep{mon03}.}
\tablenotetext{c}{Types of data presented here: I = optical spectroscopic identification;
S = time-resolved spectroscopy; T = time-series photometry; X = X-ray light curve.}
\tablenotetext{d}{Classifications are: N = novalike
variable (pulsations not confirmed); DQ = DQ Her
star or IP (evidence for pulsations); 
AM = AM Her star or polar.}
\tablenotetext{e}{Reference for period $P$, presumed to be the spin period:
(1) this paper, optical; (2) \citealt{ber15}, X-ray; (3) this paper, X-ray.}
\tablenotetext{f}{Or possibly its 1-day alias, 1390.2~s. A beat period of
1530~s is also detected.}
\label{tab:objects}
\end{deluxetable}

\begin{deluxetable}{lrrcccc}
\tablecolumns{7}
\tabletypesize{\small}
\tablewidth{0pt}
\tablecaption{Radial Velocities}
\tablehead{
\colhead{Star} &
\colhead{Time} &
\colhead{$v$} &
\colhead{$\sigma$} & 
\colhead{HA} & 
\colhead{Exposure} & 
\colhead{Telescope} \\
\colhead{} &
\colhead{} &
\colhead{(km s$^{-1}$)} &
\colhead{(km s$^{-1}$)} &
\colhead{(hh:mm)} &
\colhead{(s)} &
\colhead{} 
}
\startdata
PBC J0325.6$-$0820      &  56680.6241 & $   32$  & $    3$ &  +00:09 &  600 & H \\
PBC J0325.6$-$0820      &  56680.6314 & $   69$  & $    3$ &  +00:19 &  600 & H \\
PBC J0325.6$-$0820      &  56680.6393 & $   89$  & $    3$ &  +00:31 &  600 & H \\
PBC J0325.6$-$0820      &  56680.6466 & $   80$  & $    4$ &  +00:41 &  600 & H \\
PBC J0325.6$-$0820      &  56680.6540 & $   67$  & $    4$ &  +00:52 &  600 & H \\
PBC J0325.6$-$0820      &  56680.6845 & $ -106$  & $    4$ &  +01:36 &  600 & H \\
PBC J0325.6$-$0820      &  56680.6918 & $  -56$  & $    5$ &  +01:46 &  600 & H \\
\enddata
\tablecomments{Emission-line radial velocities.  The time given is the barycentric Julian date
of mid-integration, minus 2,440,000.0, 
on the UTC system.  The hour angle is at the start of the exposure. 
The telescope code is M for McGraw-Hill 1.3m, and H for Hiltner 2.4m.  
(This table is available in its entirety in machine-readable
and Virtual Observatory (VO) forms in the online journal. A portion is
shown here for guidance regarding its form and content.)}
\label{tab:velocities}
\end{deluxetable}

\clearpage

\begin{deluxetable}{lllrrcc}
\tablecolumns{7}
\footnotesize
\tablewidth{0pt}
\tablecaption{Fits to Radial Velocities}
\tablehead{
\colhead{Data set} & 
\colhead{$T_0$\tablenotemark{a}} & 
\colhead{$P_{\rm spec}$} &
\colhead{$K$} & 
\colhead{$\gamma$} & 
\colhead{$N$} &
\colhead{$\sigma$\tablenotemark{b}}  \\ 
\colhead{} & 
\colhead{} &
\colhead{(d)} & 
\colhead{(km s$^{-1}$)} &
\colhead{(km s$^{-1}$)} & 
\colhead{} &
\colhead{(km s$^{-1}$)}
}
\startdata
\pbcOhThreeTwoFive\tablenotemark{c} & 56685.5866(18) & 0.0871933(14) &  120(13) & $-3(10)$ & 37 &  33 \\ 
Alternate             & 56685.589(2) & 0.0856958(19) &  117(14) & $ 0(11)$ & 37 &  36 \\[1.2ex]
\swiftOhFiveOhThree & 56683.6266(9) & 0.05662(9) & 189(20) & $ 51(14)$ & 24 &  49 \\
Alternate & 56683.5779(10) & 0.05356(6) & 183(22) & $ 36(15)$ & 24 &  53 \\[1.2ex]
\pbcOhSevenOhSix      & 56681.8996(5) & 0.070907(11) &  303(13) & $-25(9)$ & 27 &  31 \\[1.2ex] 
\swiftOhNineThreeNine\tablenotemark{d} & 57107.853(4) & 0.3546(9) &  86(6) & $ 70(4)$ & 40 &  20 \\[1.2ex]
\pbcOneSevenFourOh\tablenotemark{e}  & 57195.882(4) & 0.07011(20) &  30(9) & $-34(7)$ & 113 &  30 \\[1.2ex] 
\swiftTwoOneTwoFour   & 56605.091(17) & 0.84913(11)\tablenotemark{f} &  115(12) & $-17(9)$ & 132 &  42 \\[1.2ex]
\swiftTwoThreeFourOne & 56548.7460(16) & 0.15424(20) & 151(13) & $-21(8)$ & 38 &  27 \\[1.2ex]
\enddata
\tablecomments{Parameters of least-squares fits to the radial
velocities, of the form $v(t) = \gamma + K \sin\,[2 \pi(t - T_0)/P_{\rm spec}]$.}
\tablenotetext{a}{Heliocentric Julian Date minus 2,400,000.  The epoch is chosen
to be near the center of the time interval covered by the data, and
within one cycle of an actual observation.}
\tablenotetext{b}{Root-mean-square residual of the fit.}
\tablenotetext{c}{The two periods reflect different choices of daily cycle count over
a 5-day gap, and the excessively precise periods tabulated here reflect in turn an 
arbitrary choice of cycle count between observing runs.  For each choice of daily 
cycle count, the uncertainty in the gross period is $\sim 9 \times 10^{-5}$ d.
See Section~\ref{sec:pbc0325} for details.}
\tablenotetext{d}{Fit to velocities from 2015 March only.}
\tablenotetext{e}{Velocities are from \ion{He}{1} emission; the daily cycle count is uncertain.}
\tablenotetext{f}{The daily cycle count is reasonably secure, but there is an ambiguity at the 
scale of 1 cycle per 57 days.}
\label{tab:parameters}
\end{deluxetable}

\begin{figure}
\epsscale{0.90}
\vspace{-0.5in}
\centerline{
\includegraphics[angle=0,width=1.1\linewidth,clip=]{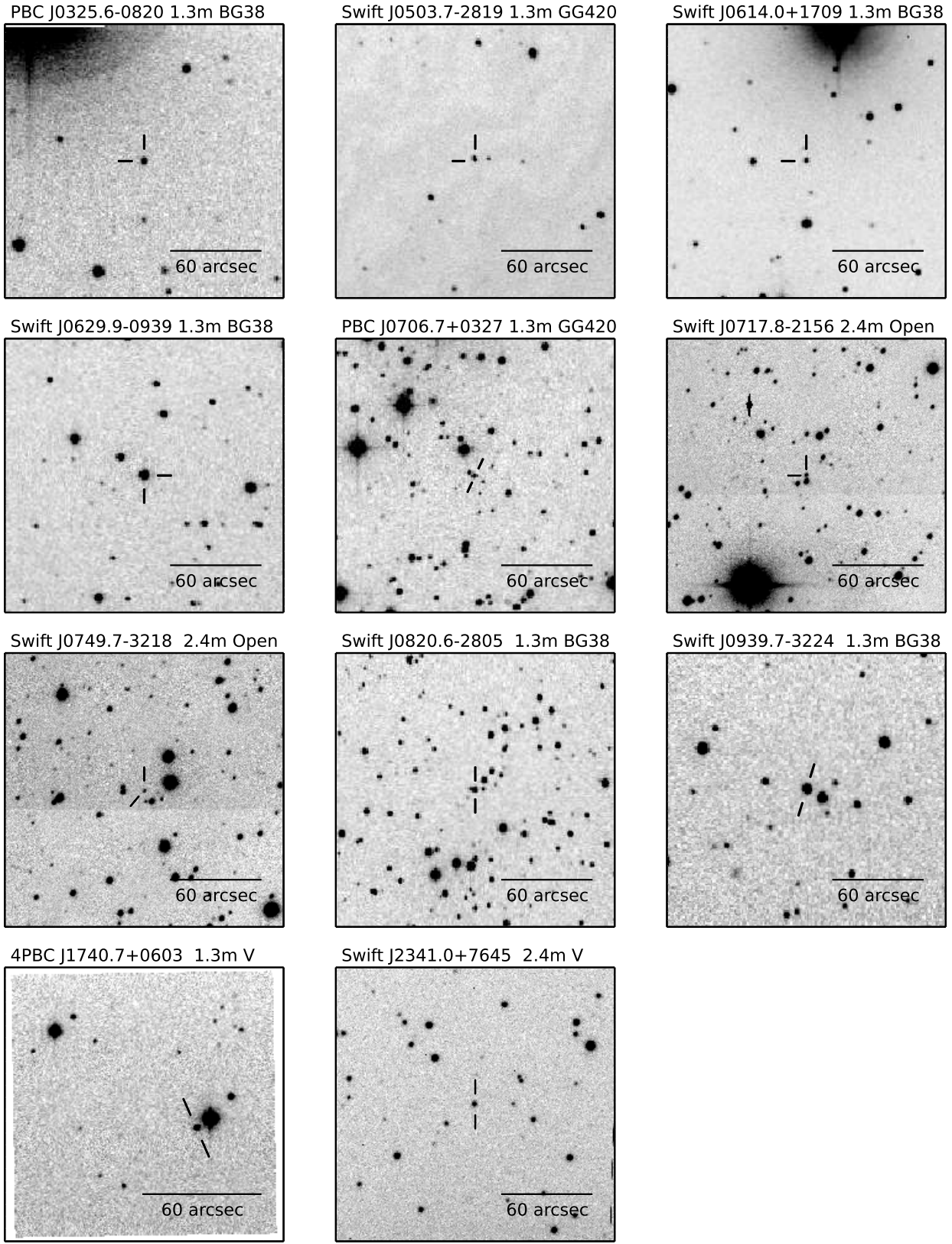}
}
\vspace{-0.5in}
\caption{Finding charts for 11 of the objects from our 1.3m or 2.4m
images.  North is up, east is to the left, and the scale is indicated.
The label on each chart gives the source of the image used, and the
tick marks are accurately aligned with the coordinates in
Table~\ref{tab:objects}.  
}
\label{fig:charts}
\end{figure}

\begin{figure}
\epsscale{1.1}
\vspace{-1.0in}
\plotone{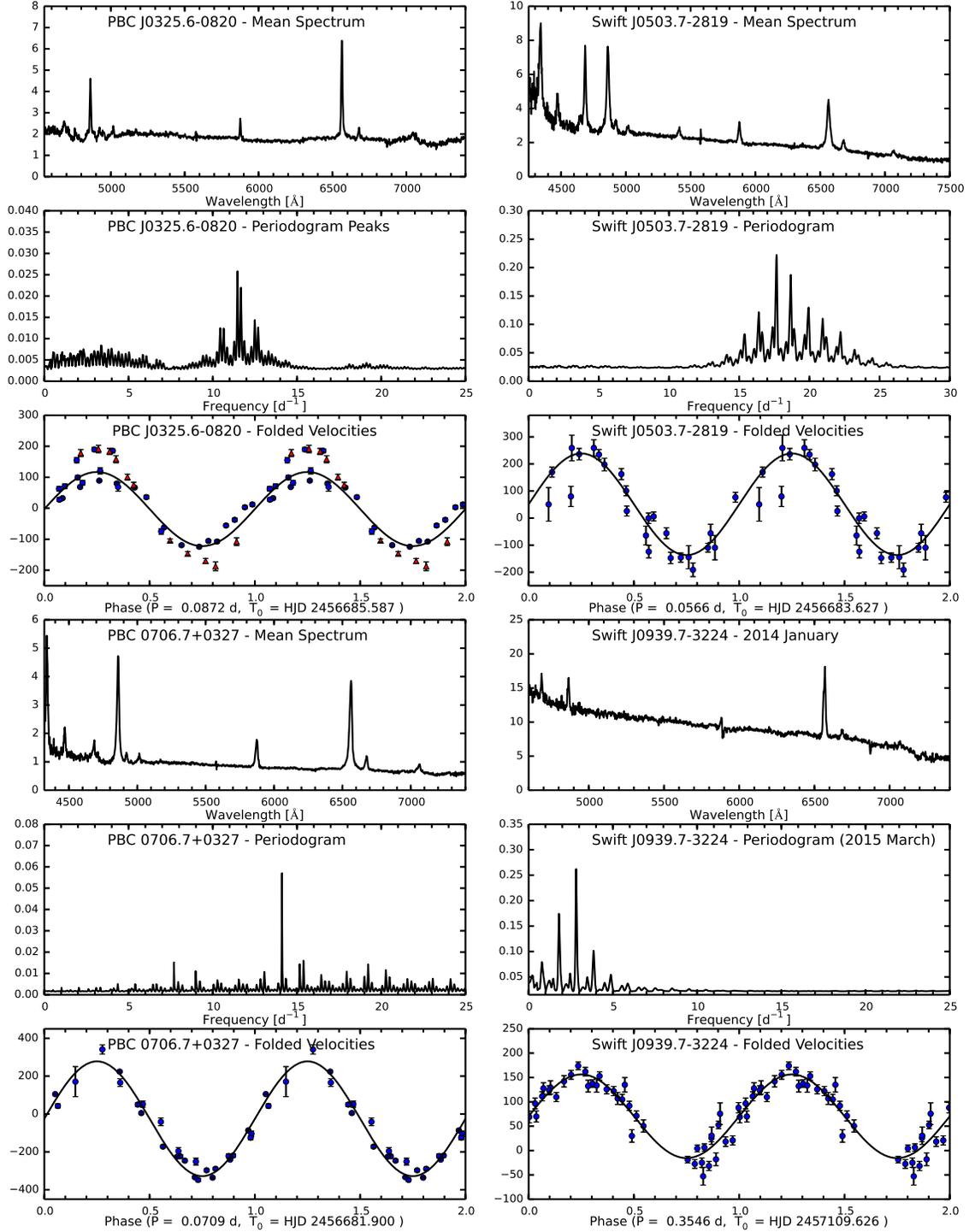}
\vspace{-0.5in}
\caption{Mean spectra, periodograms, and folded velocity curves for 
four of the objects.  The units of the vertical axes of the spectra are
$10^{-16}$ erg cm$^{-2}$ s$^{-1}$ \AA$^{-1}$; for the 
periodograms, the axis is unitless ($1 / \chi^2$); and the 
radial velocities are in km s$^{-1}$.  
In the velocity curves, all data are repeated on an extra cycle for 
continuity, the uncertainties shown are estimated from 
counting statistics, and the solid curves show the 
best-fitting sinusoids. 
For \pbcOhThreeTwoFive\ (upper left), blue circles 
show velocities from 2014 January, and red triangles from
2014 October.
}
\label{fig:montage1}
\end{figure}

\begin{figure}
\epsscale{1.1}
\vspace{-1.0in}
\plotone{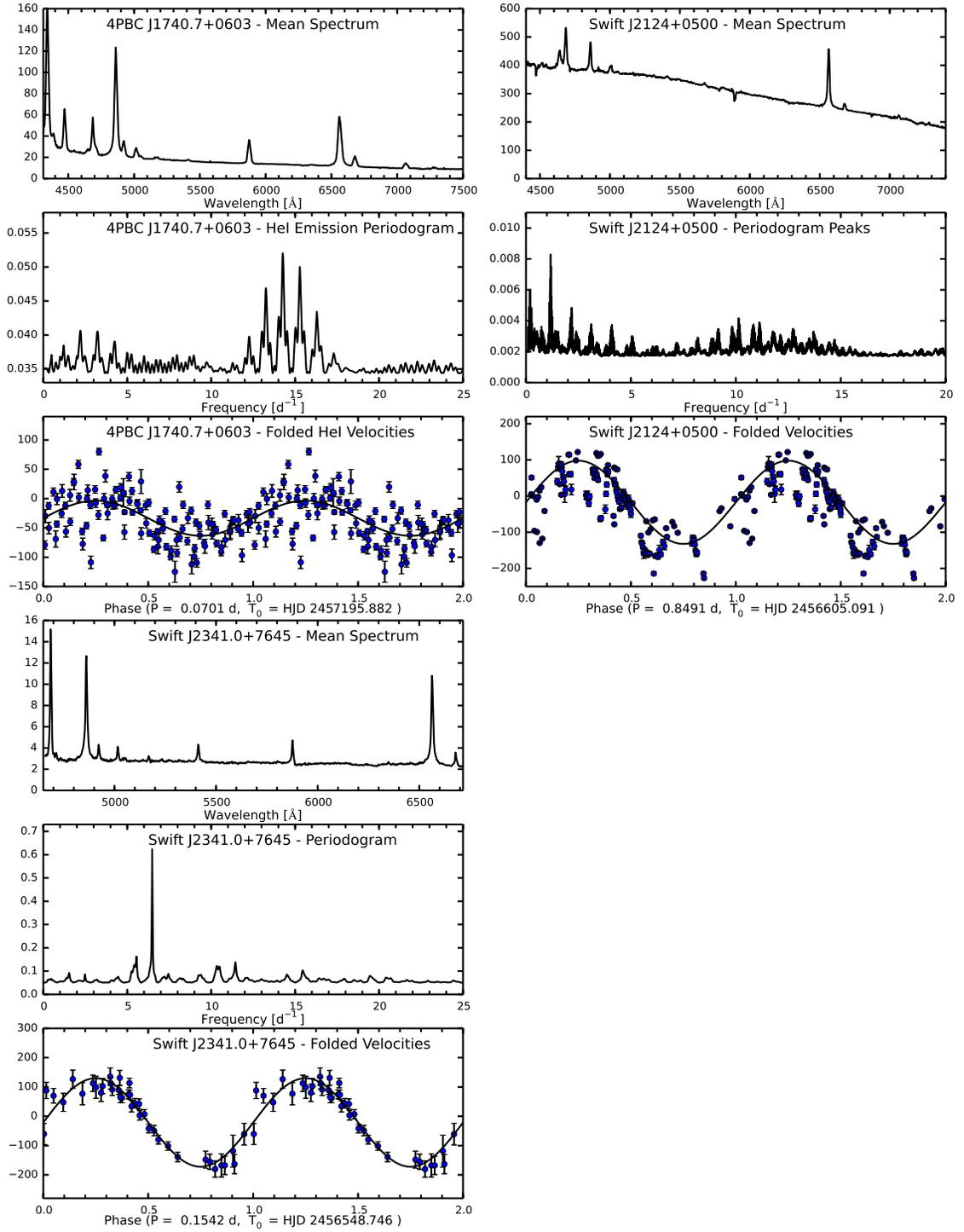}
\vspace{-0.5in}
\caption{
Same as Figure~\ref{fig:montage1}, for the remaining three objects.
}
\label{fig:montage2}
\end{figure}

\begin{figure}
\epsscale{0.9}
\vspace{-0.5in}
\plotone{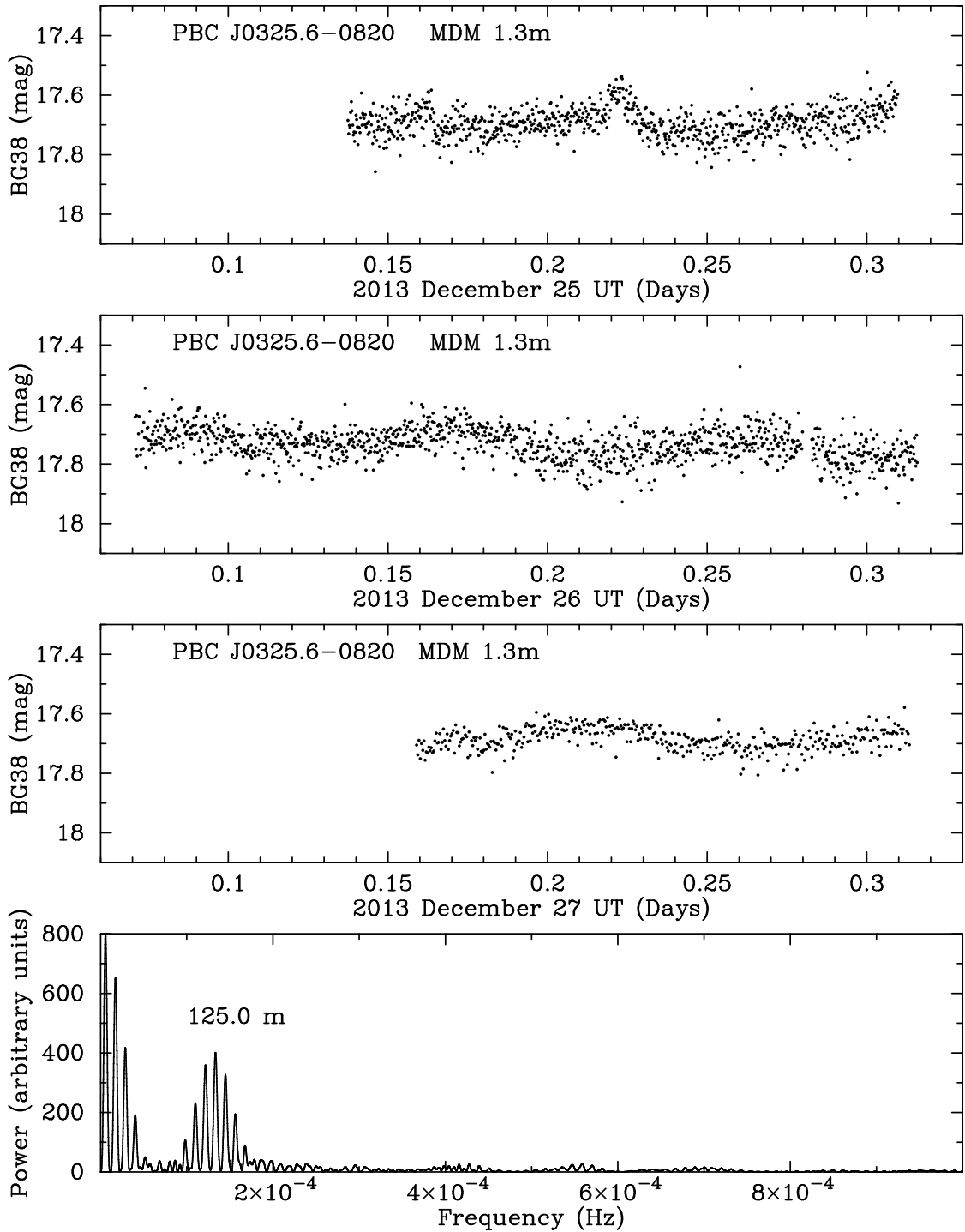}
\caption{
Time-series photometry of \pbcOhThreeTwoFive\ from the 1.3m.
Individual exposures are 15~s (December 25, 26)
or 30~s (December 27).  The highest peak at $125.0\pm0.5$~minutes
in the combined power spectrum agrees with one of the spectroscopic
candidates.
}
\label{fig:pbc0325}
\end{figure}

\begin{figure}
\epsscale{1.1}
\vspace{-1.0in}
\plotone{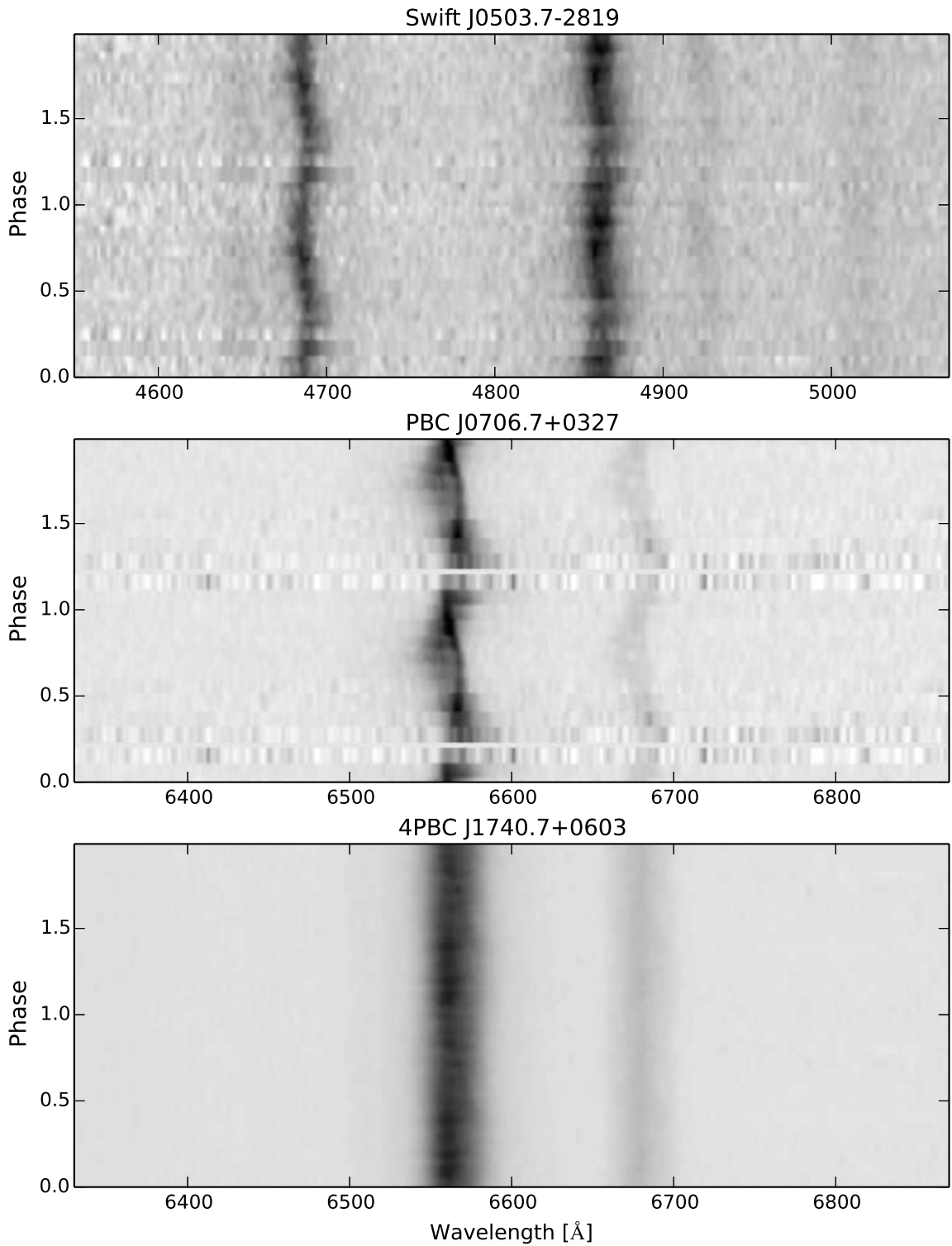}
\vspace{-0.5in}
\caption{
Portions of the spectra of three of the objects presented 
as phase-averaged greyscale
images.  The phases are computed using the periods and epochs in
Table \ref{tab:parameters}.  The images are negatives, that is,
darker represents more light.  The spectra were rectified before
phase averaging, so the continuum is forced to unity for all phases.
The horizontal features in the image for \pbcOhSevenOhSix\ 
are caused by poor coverage at the corresponding phase.  
}
\label{fig:greyscales}
\end{figure}

\begin{figure}
\epsscale{0.9}
\vspace{-1.0in}
\plotone{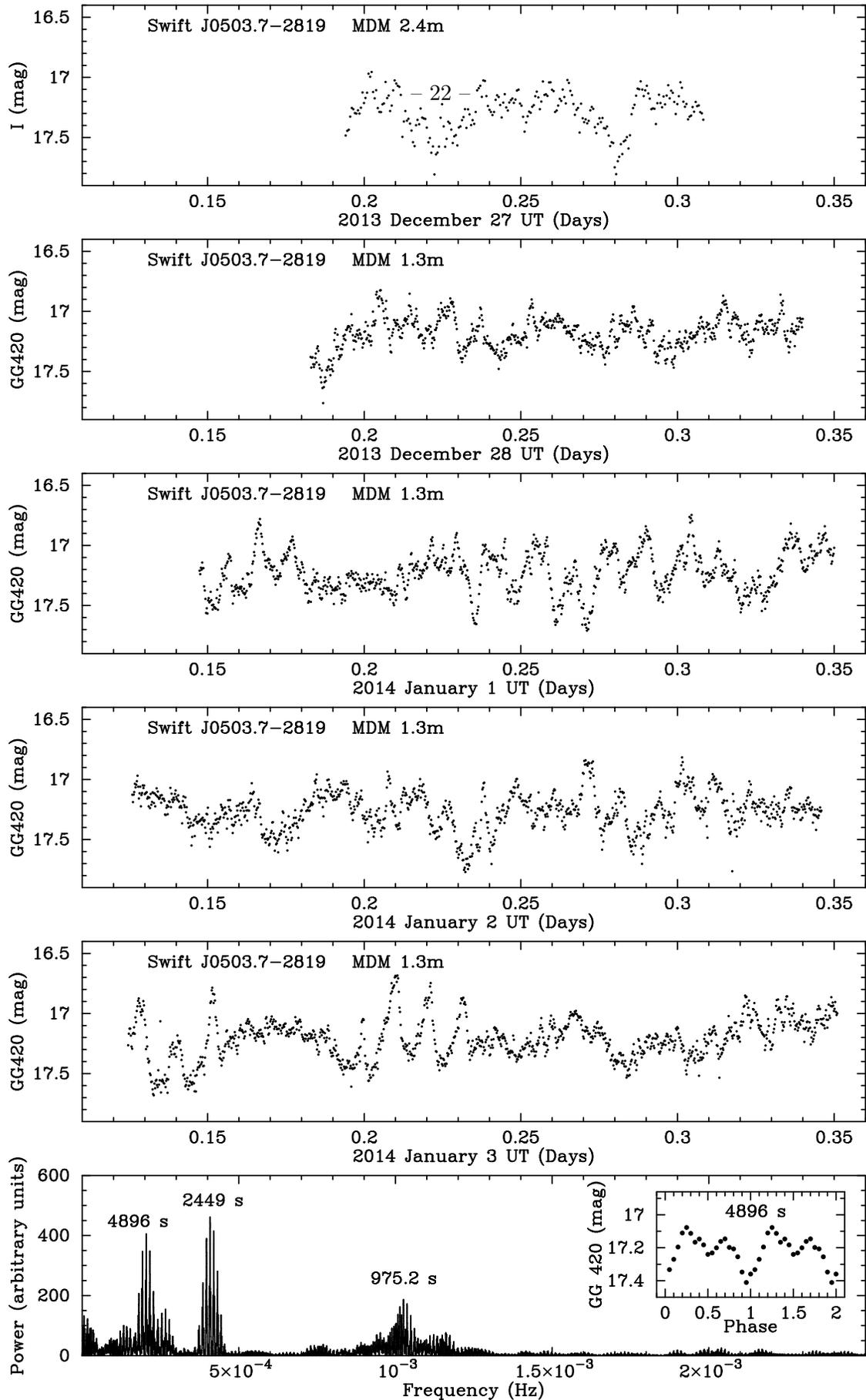}
\vspace{-0.2in}
\caption{
Time-series photometry of \swiftOhFiveOhThree.  Individual exposures
are 30~s on the 2.4m and 15~s on the 1.3m.  The power spectrum of
the combined data shows three periods, of which the longest (4896~s)
agrees with the spectroscopic orbital value, and is used to fold the
data in the inset.  The peak at 975.2~s is possibly the spin period.
}
\label{fig:swift0503}
\end{figure}

\begin{figure}
\epsscale{0.95}
\vspace{-0.9in}
\plotone{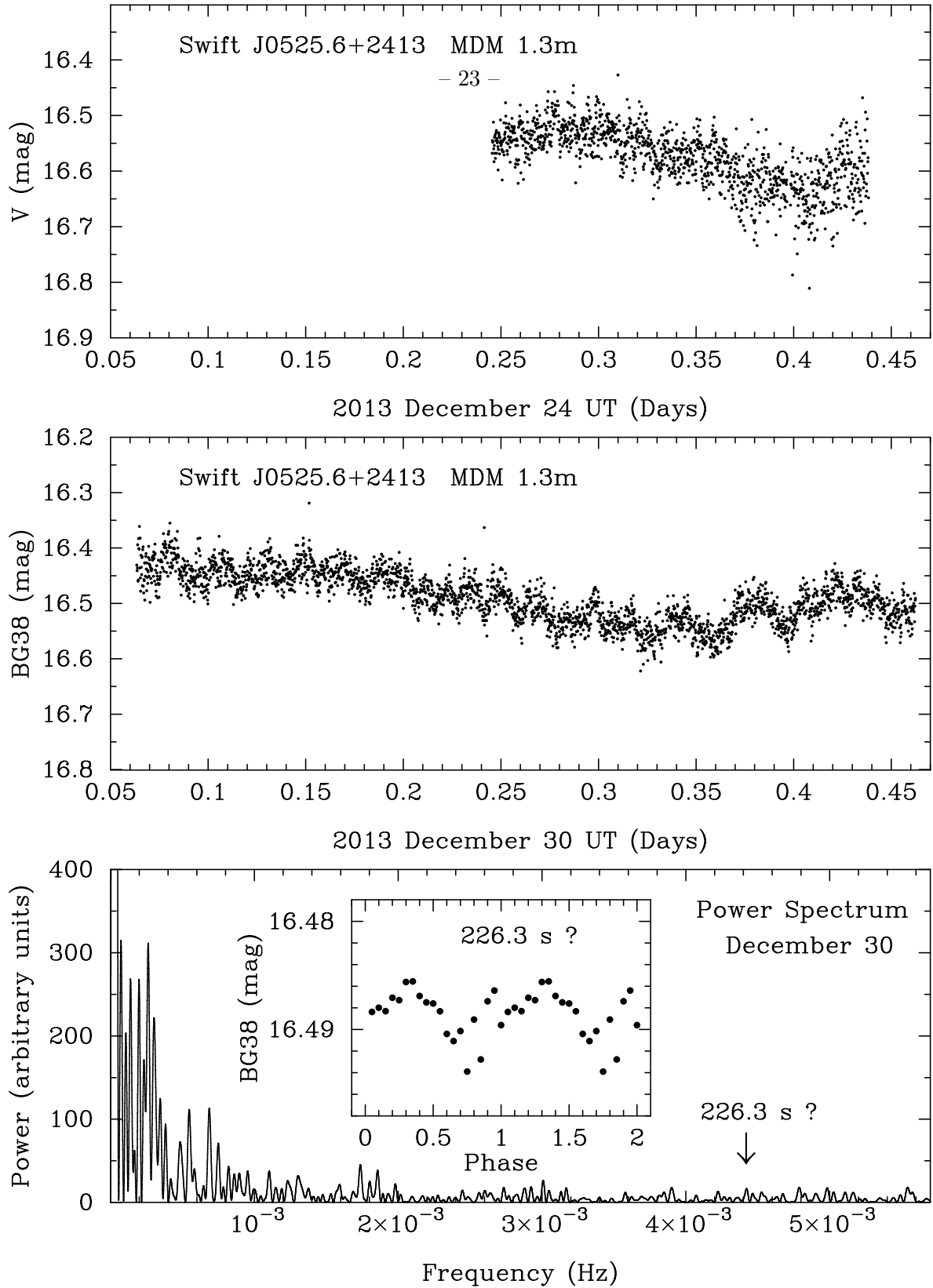}
\caption{
Time-series photometry of \swiftOhFiveTwoFive.  Individual exposures
are 10~s.  The power spectrum of the 2013 December 30 light curve
does not show a significant signal at the 226.3~s period detected
in X-rays \citep{ber15}.
Inset: a fold at 226.3~s has a half amplitude of at most 0.004~mag.
}
\label{fig:swift0525}
\end{figure}

\begin{figure}
\vspace{-0.25in}
\centerline{
\includegraphics[angle=0,width=0.6\linewidth,clip=]{fig8ab.eps}
}
\vspace{0.13in}
\centerline{
\includegraphics[angle=0,width=0.6\linewidth,clip=]{fig8cd.eps}
}
\vspace{-0.05in}
\caption{
Identification spectra of four \sw-BAT sources obtained
on the MDM 2.4m with the CCDS or OSMOS.
(a) \swiftOhSixOneFour. (b) \swiftOhSixTwoThree.
(c) \swiftOhSevenOneSeven. (d) \swiftOhSevenFourNine.
Balmer, \ion{He}{1}, and \ion{He}{2} emission
lines identify them as CVs. Fluxes are not calibrated.
}
\label{fig:spectra}
\end{figure}

\begin{figure}
\epsscale{0.8}
\vspace{-0.8in}
\plotone{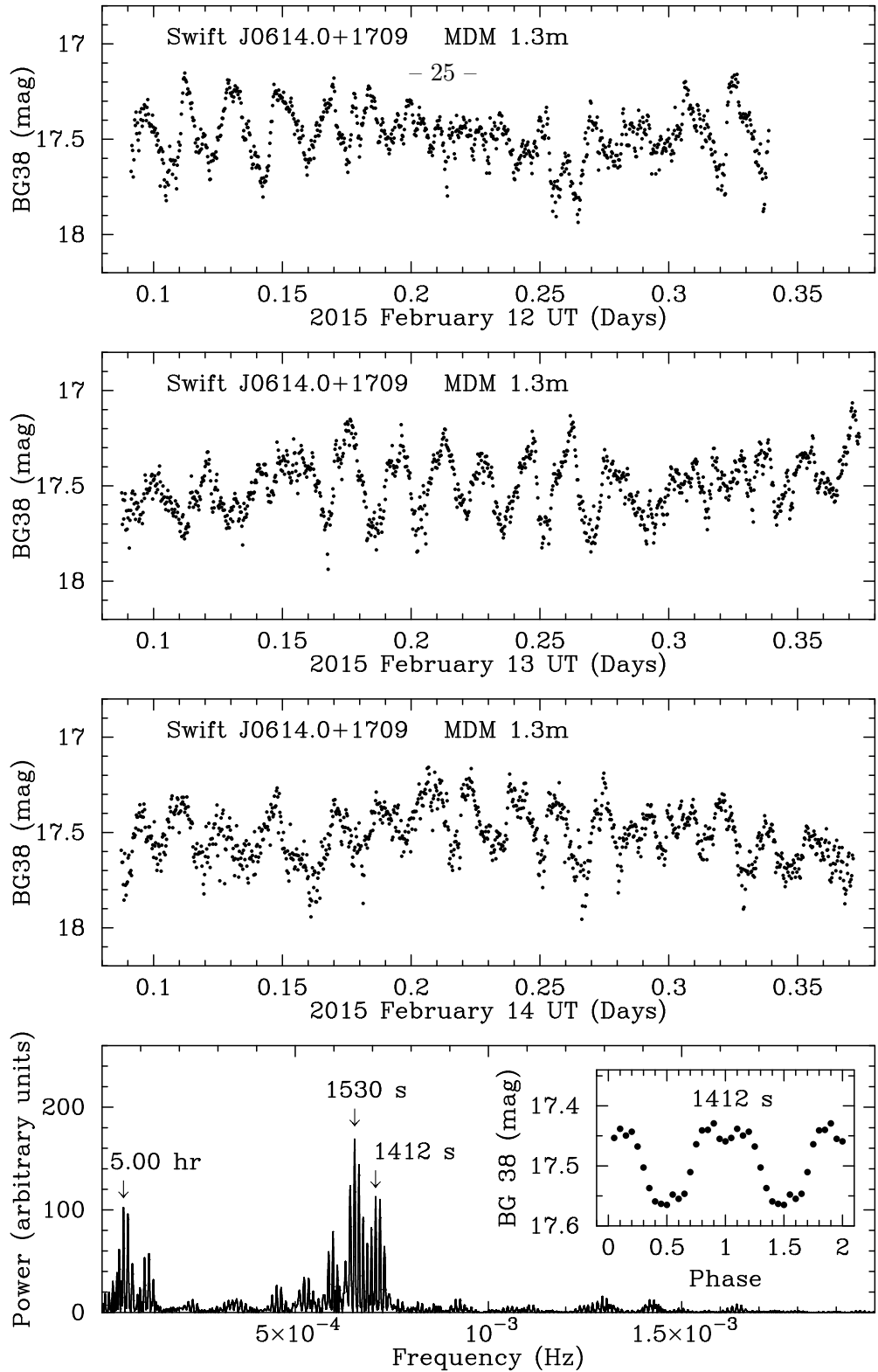}
\caption{
Time-series photometry of \swiftOhSixOneFour.  Individual exposures are 20~s.
The power spectrum of the combined data shows three periods,
of which the shortest (1412~s) is likely to be the spin period,
the longest (5.00 hr) the orbital period, and the strongest (1530~s)
the beat period between the spin and the orbit.
(One-day aliases of the spin and orbit are an alternative possibility:
1390~s and 4.15~hr.)  In the inset,
all of the data are folded on the presumed spin period.
}
\label{fig:swift0614}
\end{figure}

\begin{figure}
\epsscale{0.95}
\vspace{-1.0in}
\plotone{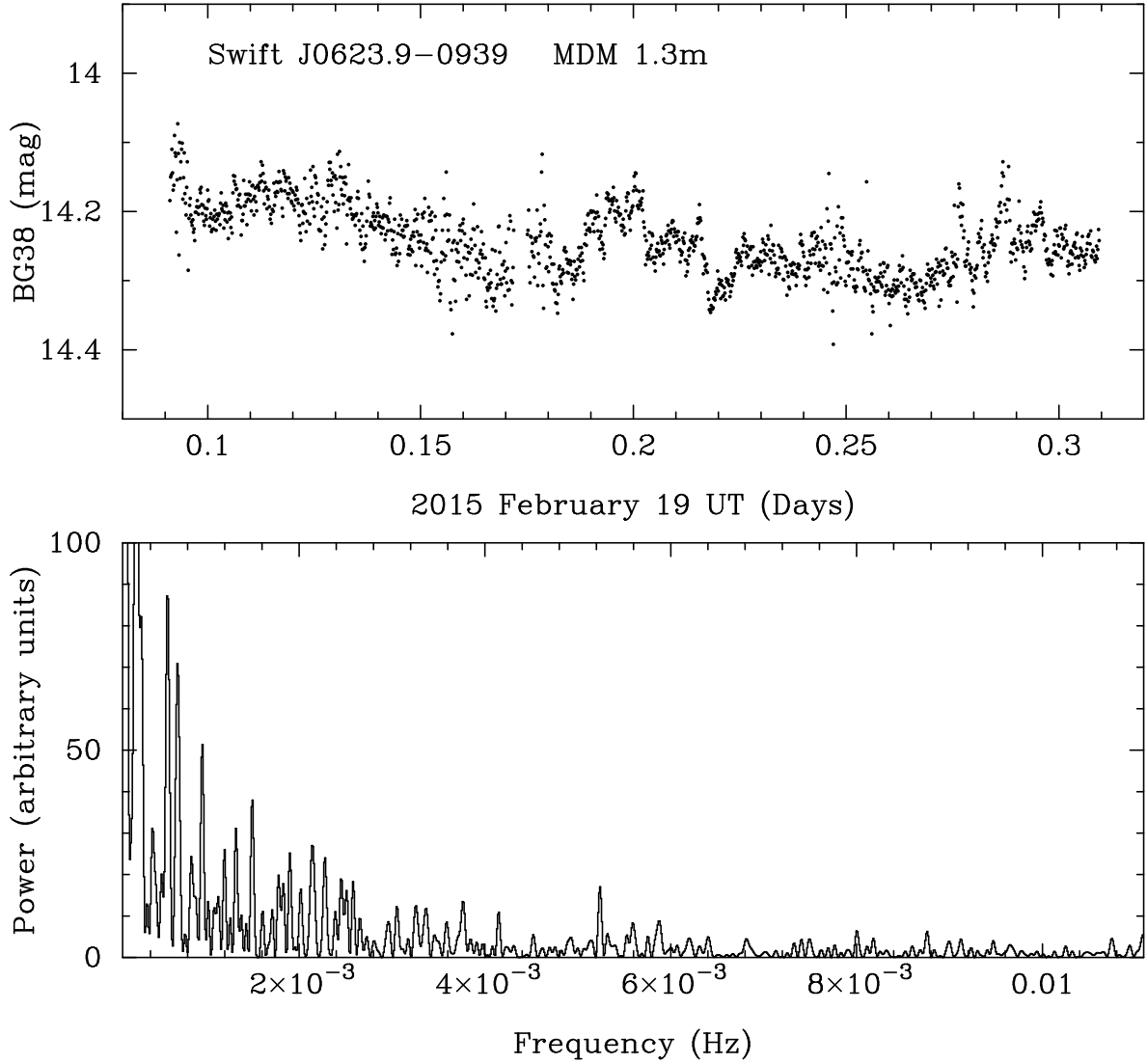}
\caption{
Time-series photometry of \swiftOhSixTwoThree.  Individual exposures
are 10~s.  This bright object was observed under partly cloudy
conditions, which is responsible for some of the scatter.
There is no period evident in the power spectrum.
}
\label{fig:swift0623}
\end{figure}

\begin{figure}
\epsscale{0.95}
\vspace{-1.0in}
\plotone{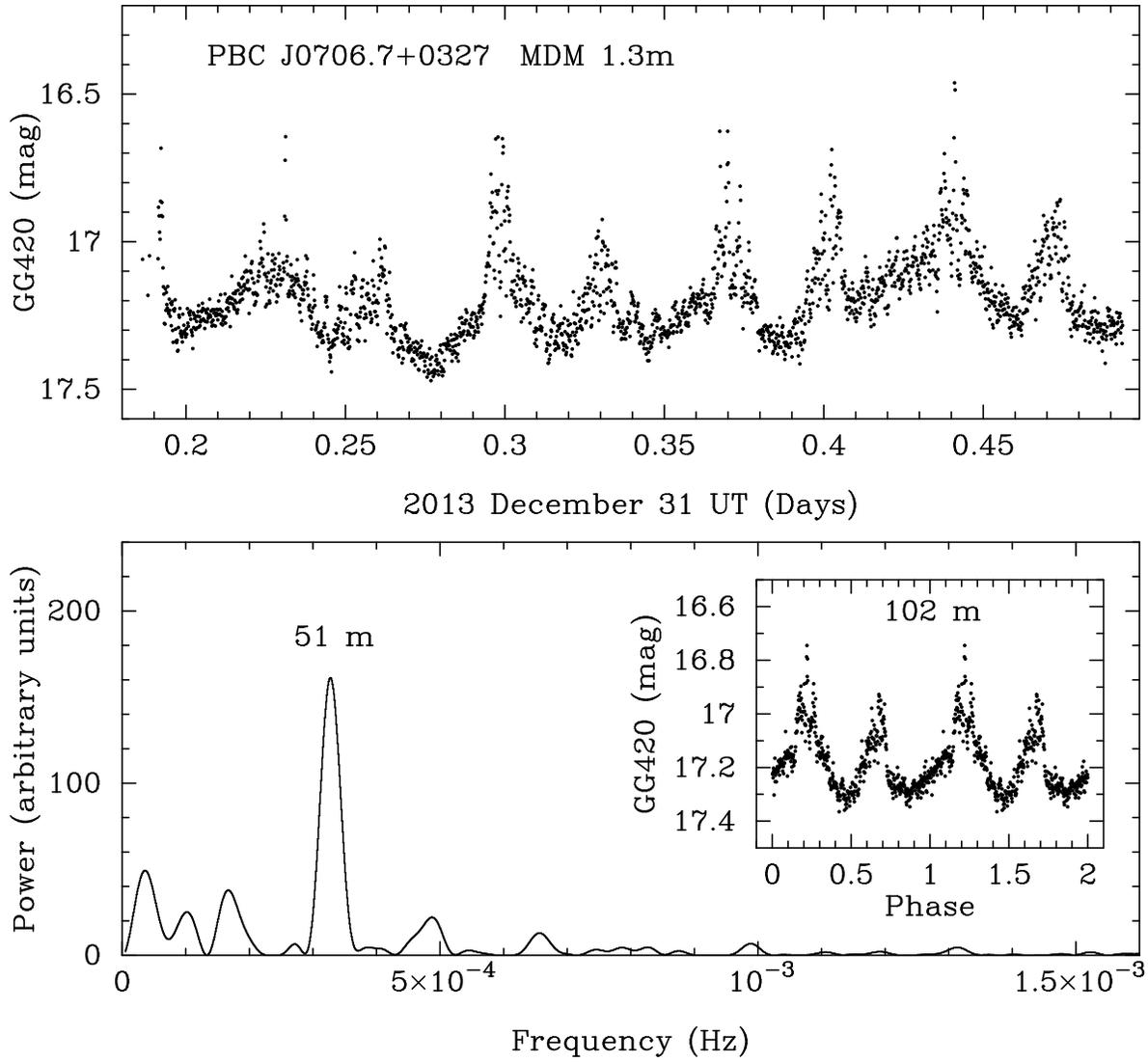}
\caption{
Time-series photometry of \pbcOhSevenOhSix\ from the 1.3m.
Individual exposures are 10~s.  The peak in the power
spectrum at 51 minutes corresponds to half the spectroscopic
orbital period.  The inset shows the mean double-peaked light
curve folded on the 102 minute orbital period.
}
\label{fig:pbc0706}
\end{figure}

\begin{figure}
\epsscale{0.95}
\vspace{-1.0in}
\plotone{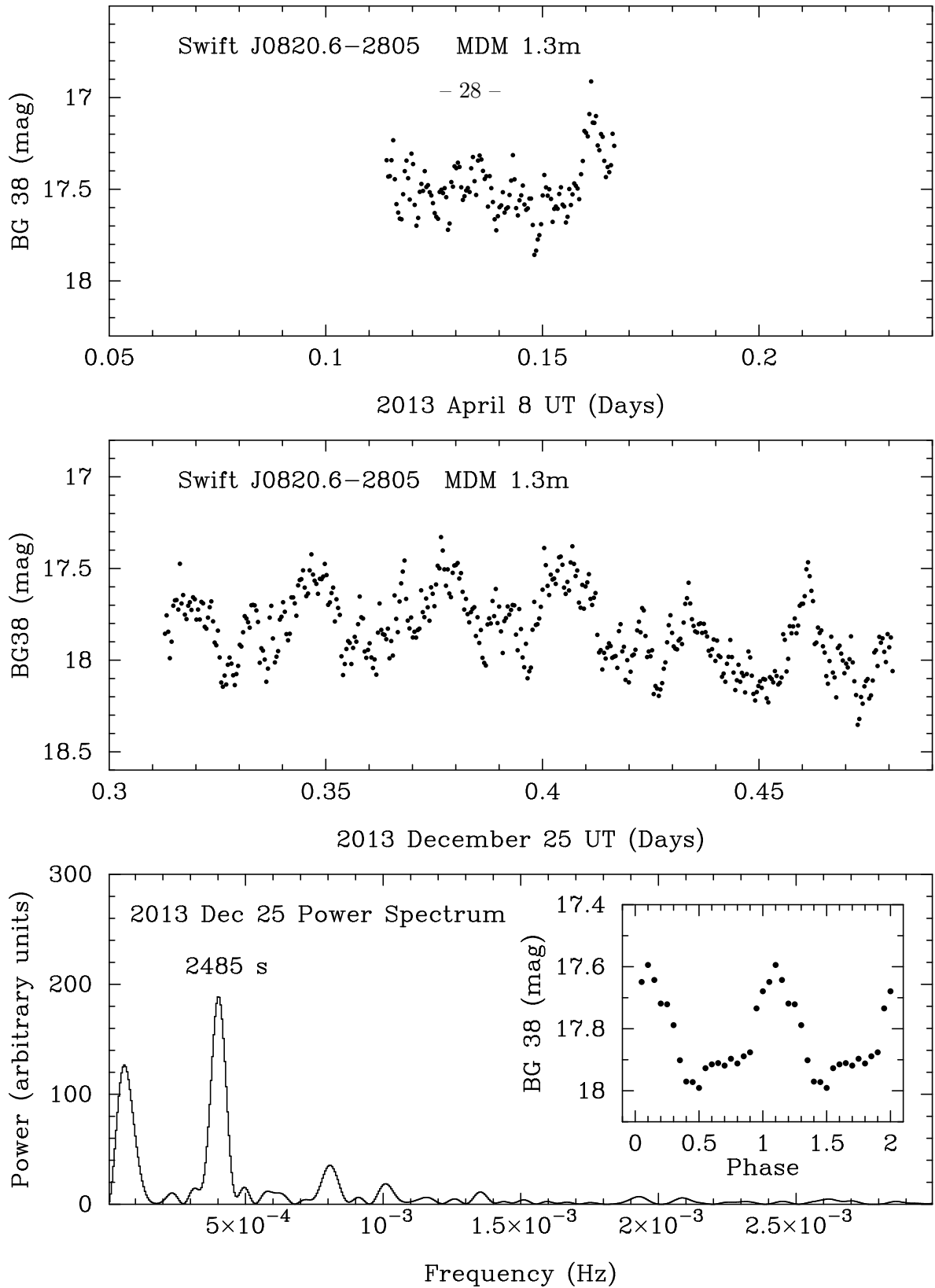}
\caption{
Time-series photometry of \swiftOhEightTwoOh\ from the 1.3m.
Individual exposures are 30~s.  The inset shows the mean
light curve from 2013 December 25 folded on the tentative
$2485\pm50$~s period.
}
\label{fig:swift0820}
\end{figure}

\begin{figure}
\epsscale{0.9}
\vspace{-1.0in}
\plotone{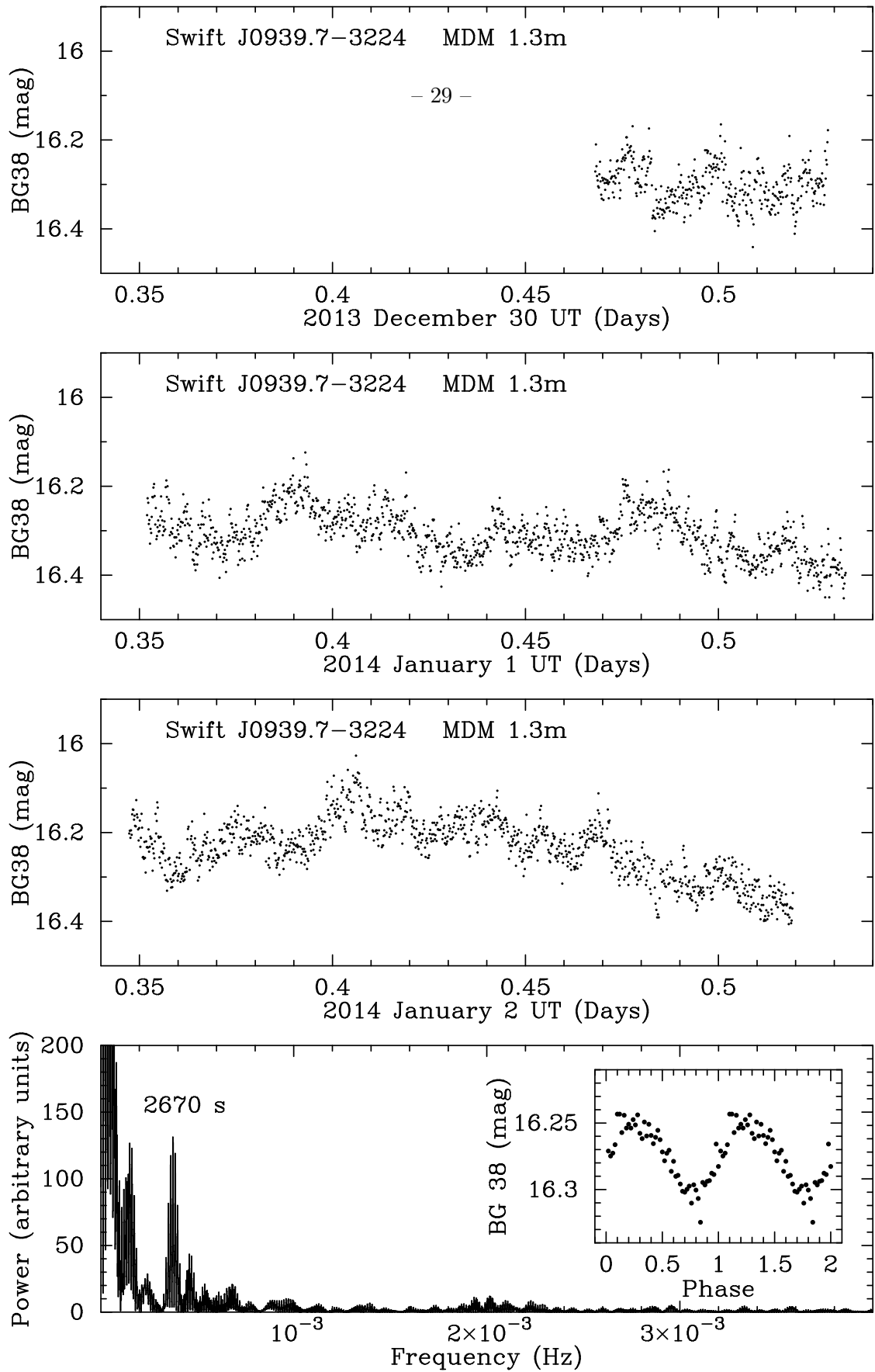}
\caption{
Time-series photometry of \swiftOhNineThreeNine\ from the 1.3m.
Individual exposures are 10~s.  The inset shows the mean
light curve folded on the tentative $2670\pm7$~s period.
}
\label{fig:swift0939}
\end{figure}

\begin{figure}
\epsscale{0.95}
\vspace{-1.0in}
\plotone{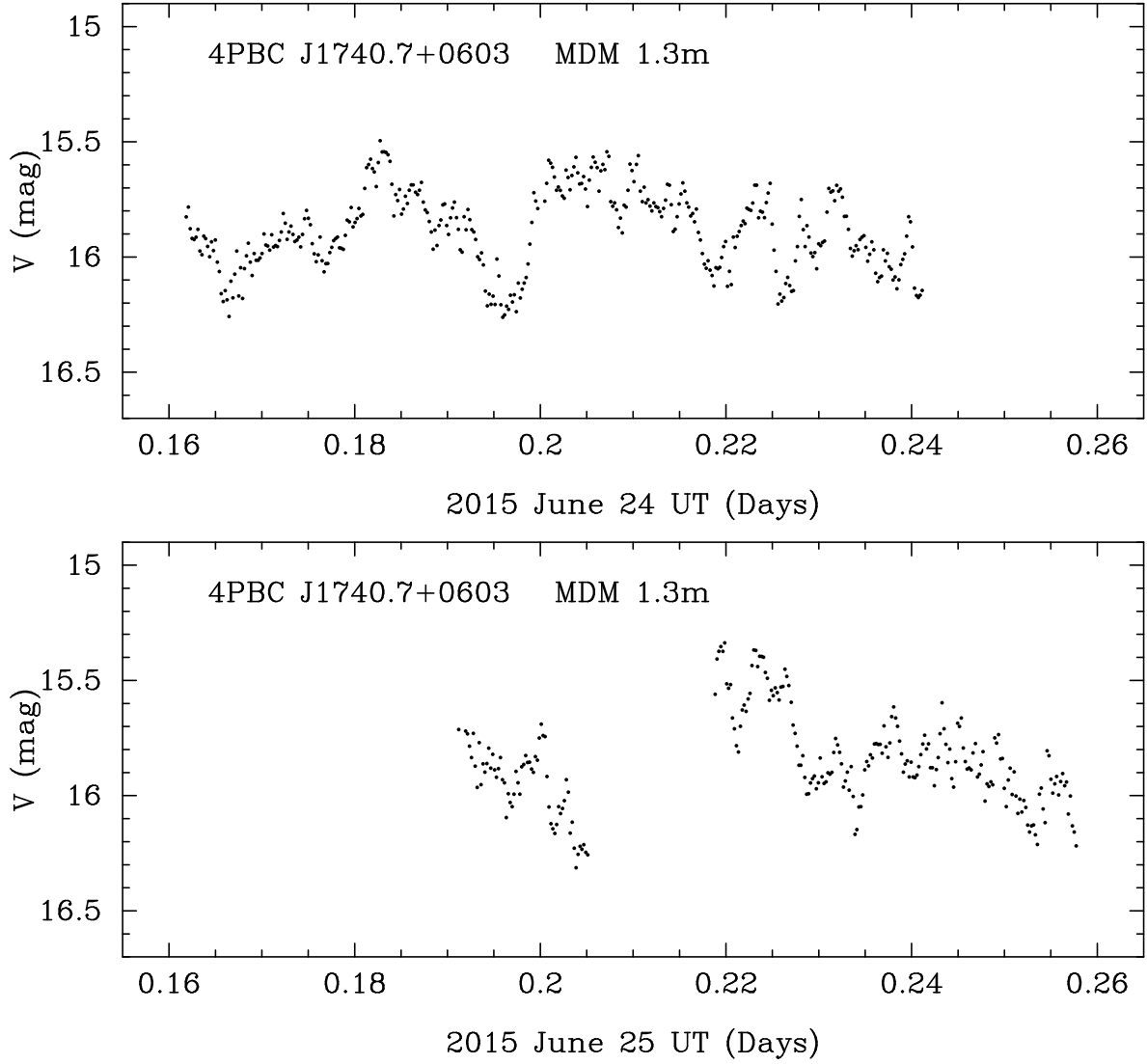}
\caption{
Time-series photometry of \pbcOneSevenFourOh\ in the $V$ filter
from the 1.3m.  Individual exposures are 15~s.  The magnitude
is calibrated with respect to the adjacent bright star
seen in Figure~\ref{fig:charts}, Tyc 427-1298-1
which has $V=11.53\pm0.12$ \citep{hog00}.
}
\label{fig:pbc1740}
\end{figure}

\begin{figure}
\epsscale{0.9}
\vspace{-1.0in}
\plotone{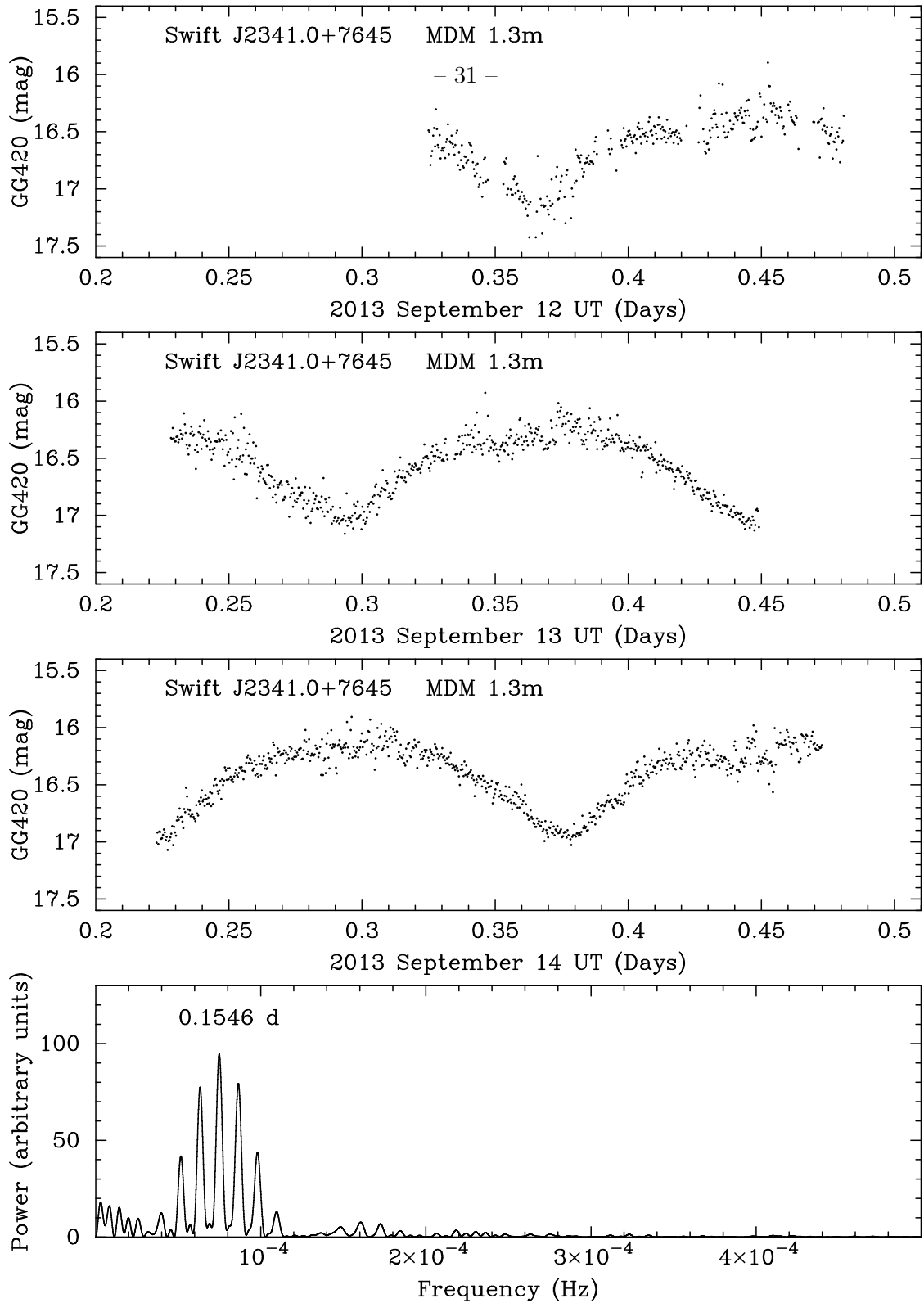}
\caption{
Time-series photometry of \swiftTwoThreeFourOne\ using the Andor
camera on the 1.3m.  Individual exposures are 30~s.  The photometric
period in the power spectrum, 0.1546(6) days, is in agreement with
the spectroscopic period.
}
\label{fig:swift2341}
\end{figure}

\begin{figure}
\epsscale{0.6}
\vspace{-1.0in}
\includegraphics[angle=270,width=1.0\linewidth,clip=]{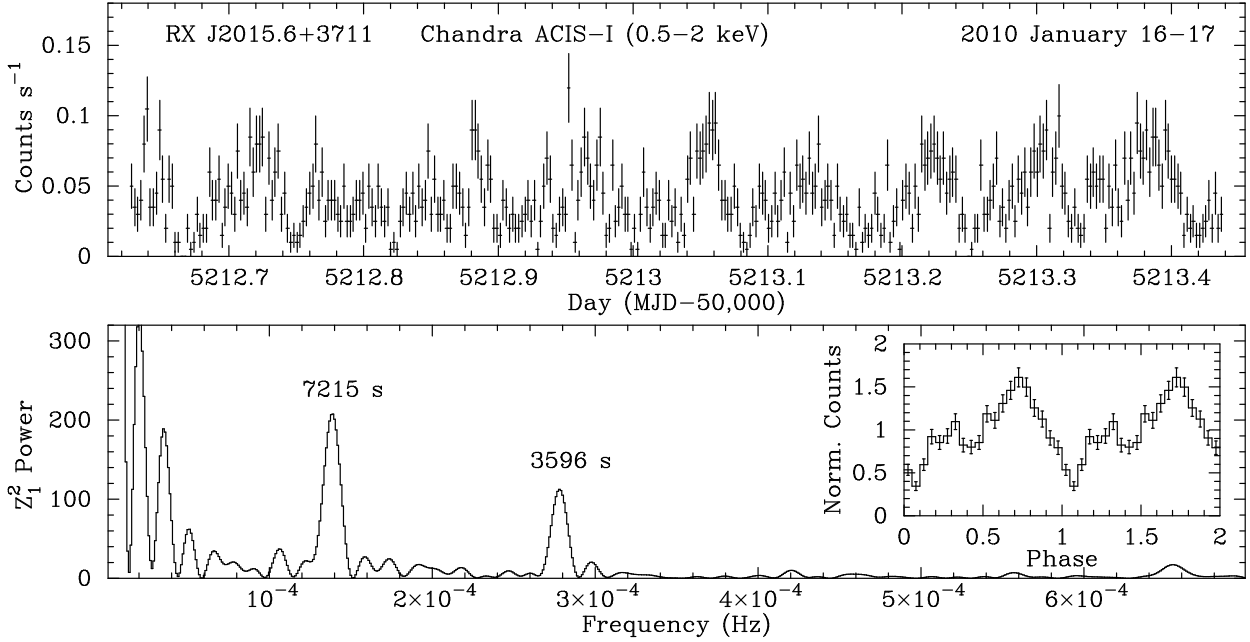}
\caption{
\chandra\ (ObsID 11092) light curve of \rxj\ in soft X-rays. 
Top: The 0.5--2~keV X-rays in 200~s bins.
Bottom: A $Z_1^2$ periodogram (Rayleigh test) of the 0.5--2~keV X-rays,
revealing the orbital period of 7215(31)~s and its harmonic at 3596(9)~s.
The inset is the background subtracted and normalized 0.5--2~keV
light curve folded at the orbital period.
See Figure~\ref{fig:rxjfold} for other energy ranges.
}
\label{fig:rxjperiod}
\end{figure}

\begin{figure}
\epsscale{0.5}
\plotone{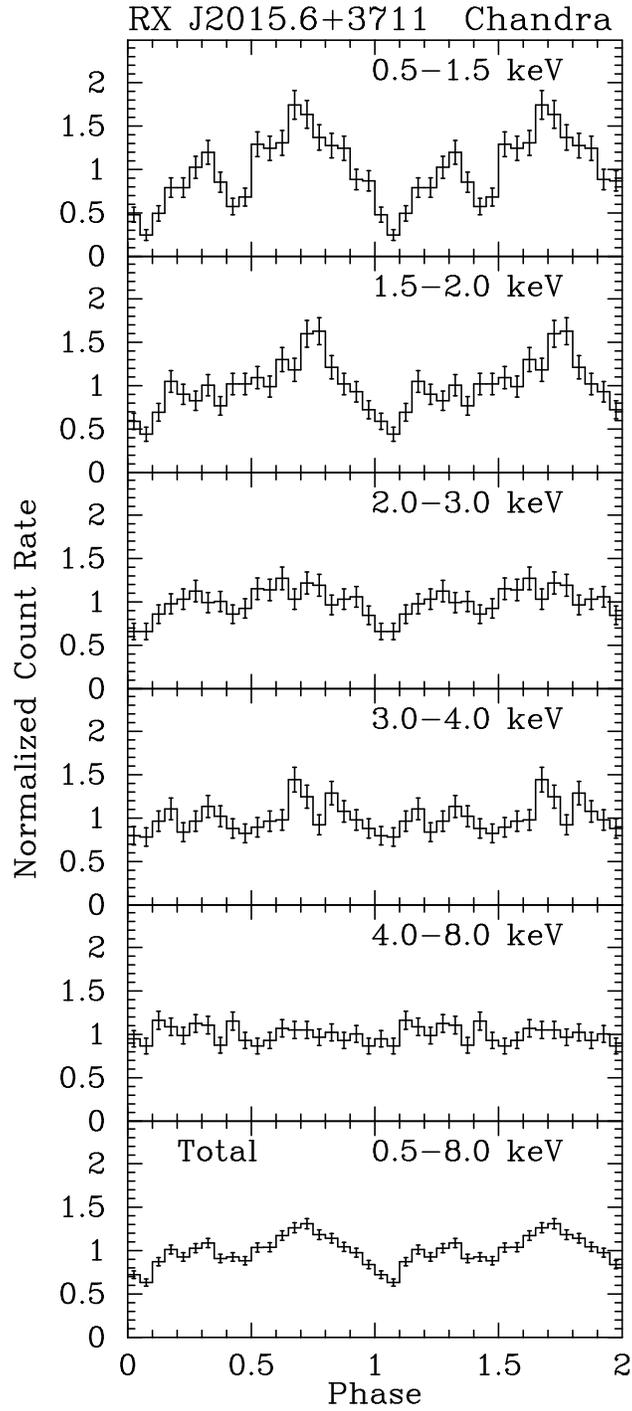}
\caption{
Energy-dependent folded light curves from the \chandra\ observation
of \rxj.  As in Figure~\ref{fig:rxjperiod}, background from
a nearby region in the image has been subtracted, and the counts
per bin are normalized to 1.
}
\label{fig:rxjfold}
\end{figure}

\end{document}